\documentclass[]{spie}  
\usepackage[]{graphicx}
\usepackage{amsmath,amssymb,url}
\title{Apodized vortex coronagraph designs for segmented aperture telescopes} 

\author{Garreth~Ruane\supit{a}, Jeffrey~Jewell\supit{b}, Dimitri~Mawet\supit{a,b}, Laurent~Pueyo\supit{c}, and Stuart~Shaklan\supit{b}
\skiplinehalf
\supit{a}California Institute of Technology, 1200 E. California Blvd., Pasadena, CA 91125, USA; \\
\supit{b}Jet Propulsion Laboratory, California Institute of Technology, 4800 Oak Grove Dr., Pasadena, CA 91109, USA;\\
\supit{c}Space Telescope Science Institute, 3700 San Martin Drive, Baltimore, MD, 21218, USA\\
}

\authorinfo{Further author information: send correspondence to gruane@astro.caltech.edu}

\graphicspath{{Figures/}}
 
\begin{document} 
  \maketitle 

\begin{abstract}
Current state-of-the-art high contrast imaging instruments take advantage of a number of elegant coronagraph designs to suppress starlight and image nearby faint objects, such as exoplanets and circumstellar disks. The ideal performance and complexity of the optical systems depends strongly on the shape of the telescope aperture. Unfortunately, large primary mirrors tend to be segmented and have various obstructions, which limit the performance of most conventional coronagraph designs. We present a new family of vortex coronagraphs with numerically-optimized gray-scale apodizers that provide the sensitivity needed to directly image faint exoplanets with large, segmented aperture telescopes, including the Thirty Meter Telescope (TMT) as well as potential next-generation space telescopes. 
\end{abstract}


\keywords{High contrast imaging, instrumentation, exoplanets, direct detection, coronagraphs}

\section{INTRODUCTION}
\label{sec:intro}  

The development of extreme adaptive optics and high-contrast imaging techniques for ground-based telescopes (e.g. GPI \cite{GPI2006}, SPHERE \cite{SPHERE2008}, and SCExAO \cite{Martinache2009}) has enabled the detection and characterization of several young, giant exoplanets \cite{Marois2008,Lafreniere2008,Lagrange2009,Macintosh2015}. However, planets within the detection limits of current instruments are relatively rare \cite{Bowler2016}. The next generation of ground-\cite{Macintosh2006,Kasper2010} and space-based \cite{Noecker2016,Feinberg2014,Dalcanton2015} telescopes will be capable of detecting fainter, older, less massive planets at smaller angular separation from their host stars, thereby providing access to planet populations with significantly higher occurrence rates. Additionally, thorough spectral characterization will be possible for many of these targets thanks to rapid technological developments for precise control and calibration of unwanted stellar radiation, including dedicated coronagraphs for diffraction suppression, wavefront control, as well as new observing and post-processing strategies. 

Detection and characterization of faint planets requires an optical system that isolates the light from the planet from noise associated with starlight. A coronagraph accomplishes this by manipulating the amplitude and phase of the incoming light such that the diffracted starlight is suppressed or removed optically prior to detection. Several elegant coronagraph designs exist that provide various levels of suppression and planet throughput \cite{Kuchner2002,Kasdin2003,Codona2004,Mawet2005,Foo2005,Guyon2005,Soummer2005,Trauger2007,Guyon2010}. The performance and complexity of each depends on the shape of the telescope pupil. Large, segmented apertures present a unique challenge; the coronagraph masks must be designed to account for the diffraction owing to discontinuities in the aperture including the secondary mirror, spider support structures, and gaps between mirror segments.
\cite{Mawet2011_improved,Mawet2013_ringapod,Carlotti2014,Ruane2015_SPIE,Ruane2015_LPM,Pueyo2013,Mazoyer2015,Guyon2014,Balasubramanian2015,Trauger2016,Zimmerman2016} (Also see Zimmerman et al. and Guyon et. al., these proceedings). 

The vortex coronagraph (VC) \cite{Mawet2005,Foo2005} has been demonstrated to provide high sensitivity to planets at small angular separations \cite{Serabyn2010}. However, complicated aperture shapes limit the performance of the conventional VCs \cite{Mawet2010b} and thereby drive the complexity of the optical design \cite{Mawet2011_improved,Mawet2013_ringapod,Carlotti2014,Ruane2015_SPIE,Ruane2015_LPM} and/or requirements for wavefront control \cite{Pueyo2013,Mazoyer2015}. This work overcomes this technical challenge by introducing a gray-scale apodizer to the VC that acts to suppress polychromatic, diffracted starlight at angular separations $<10\lambda/D$ potentially down to the $10^{-10}$ level on a segmented aperture telescope similar to those proposed for a future LUVOIR flagship mission \cite{Dalcanton2015,SCDA}. 
\newpage

\section{CORONAGRAPH OPTIMIZATION} \label{sec:opt} 

\begin{figure}[t]
    \centering
    \includegraphics[width=0.85\linewidth]{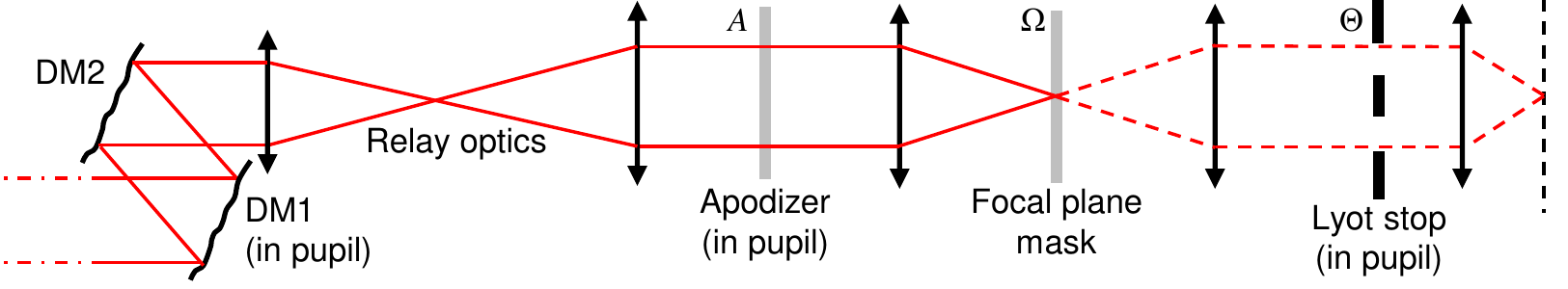}
    \caption{Schematic of a coronagraph instrument, with deformable mirrors DM1 and DM2, a pupil-plane apodizer~$A$, focal plane mask~$\Omega$, and a Lyot stop~$\Theta$ in the downstream pupil. The black arrows represent powered optics.}
    \label{fig:diagram}
\end{figure}

\begin{figure}[b]
    \centering
    \hspace{1.25cm}
    \includegraphics[width=0.7\linewidth]{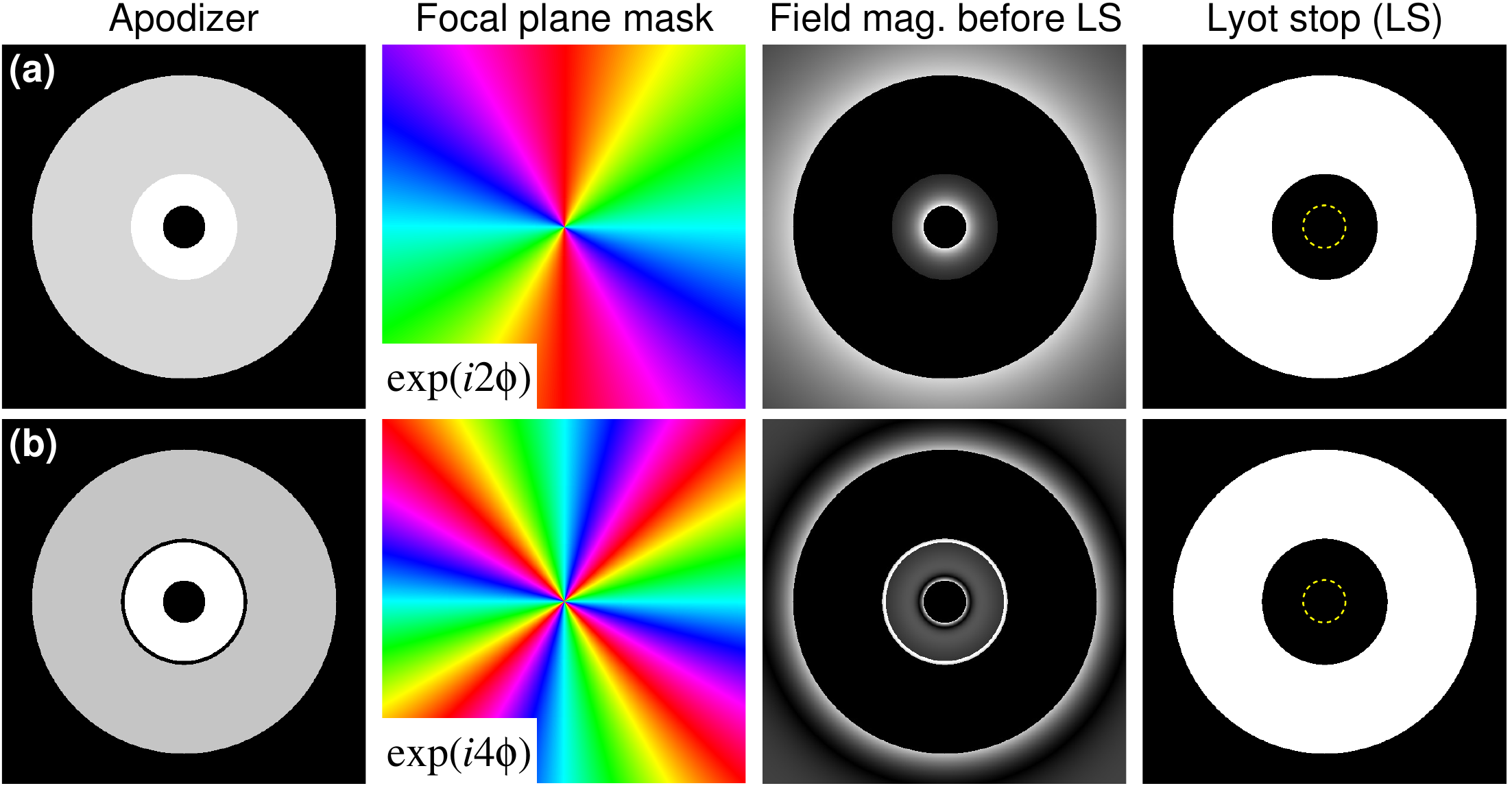}\hspace{0.2cm}
    \includegraphics[trim={0 -10mm 0 0},clip,scale=0.7]{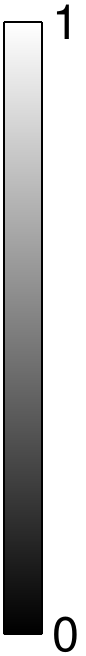}\hspace{0.2cm}
    \includegraphics[trim={0 -10mm 0 0},clip,scale=0.7]{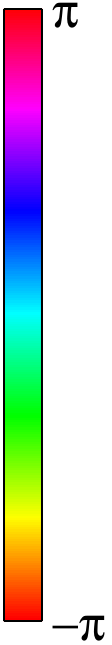}
    \caption{(a) Charge 2 and (b) charge 4 ring-apodized vortex coronagraphs designed for an annular pupil with central obscuration ratio $R_0=0.14$. The dotted line in the Lyot stop (LS) indicates the boundary of the full geometric pupil.}
    \label{fig:baseline}
\end{figure}

\subsection{The optical system and baseline design}

A high-contrast imaging coronagraph instrument is made up of a wavefront control sub-system with one or two deformable mirrors (DMs) and a series of coronagraphic masks arranged between powered optics. Fig. \ref{fig:diagram} shows an example system with two DMs, a pupil plane apodizer $A$, a focal plane mask $\Omega$, and a Lyot stop $\Theta$ in the downstream pupil. In general, a coronagraph may also include masks or surfaces displaced from the pupil and focal planes and/or in additional optical relays.

The designs presented here are based on the ring-apodized vortex coronagraph (RAVC) \cite{Mawet2013_ringapod}, which provides theoretically ideal and achromatic on-axis starlight cancellation with an annular aperture. The analytically-inspired design is shown in Fig. \ref{fig:baseline}. The apodizer is a semi-transparent ring with amplitude transmittance $t$ that extends from $R_\mathrm{gray}$ to $R$, where $R_\mathrm{gray}$ is the inner radius of the gray ring and $R$ is the outer radius of the pupil. The focal plane mask has complex transmittance $\exp(il\phi)$, where $l$ is an even, nonzero integer known as the charge and $\phi$ is the azimuthal angle. The Lyot stop is a annulus with inner and outer radii $R_\mathrm{gray}$ and $R$, assuming there is no magnification between the pupils. 

For $l=2$, the transmittance of the gray ring is given by $t = 1- (R_0/R_\mathrm{gray})^2$, where $R_0$ is the ratio of the inner and outer radii of the full annular aperture. For $l=4$, there is an additional narrow black ring from $R_\mathrm{black}$ to $R_\mathrm{gray}$, $R_\mathrm{gray} = \sqrt{(R_\mathrm{black}^4-R_0^4)/(R_\mathrm{black}^2-R_0^2)}$, and the transmittance of the gray ring is given by $t = (R_\mathrm{black}^2-R_0^2)/R_\mathrm{gray}^2$. The remaining free parameter is varied to maximize the throughput. 

Several coronagraph throughput definitions may be found in the literature. In this work, we use two measures: the total energy throughput (i.e. the fraction of energy from a point source that transmits through the Lyot stop) and the fraction of energy within 0.7$\lambda/D$ of the source position, which is roughly the half-width half-maximum of an ideal point spread function (PSF). The latter definition is the relevant quantity for detecting point sources in noisy data using aperture photometry, as is the case in common post-processing approaches for exoplanet detection. We note, however, that this may be a conservative estimate for throughput provided advanced matched-filtering and local deconvolution techniques can make use of the planet light outside of the PSF core, which will be the topic of future studies.  

\begin{figure}
    \centering
    \includegraphics[height=0.322\linewidth]{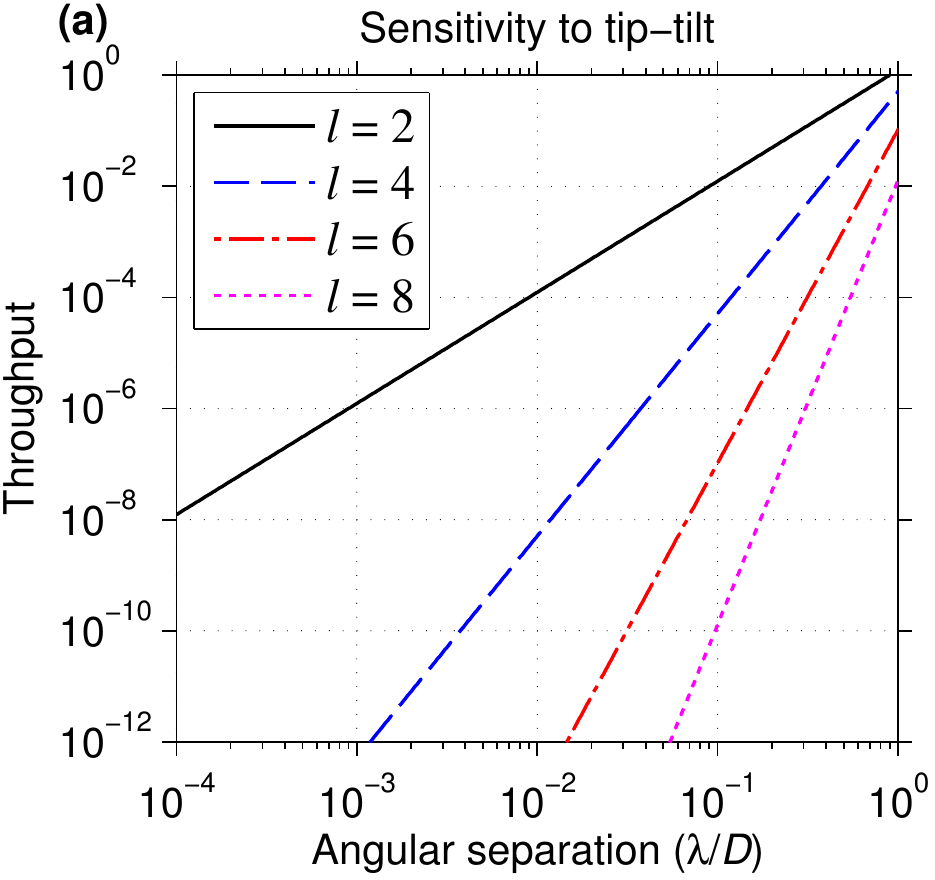}
    \includegraphics[trim={0.6cm 0 0 0},clip,height=0.322\linewidth]{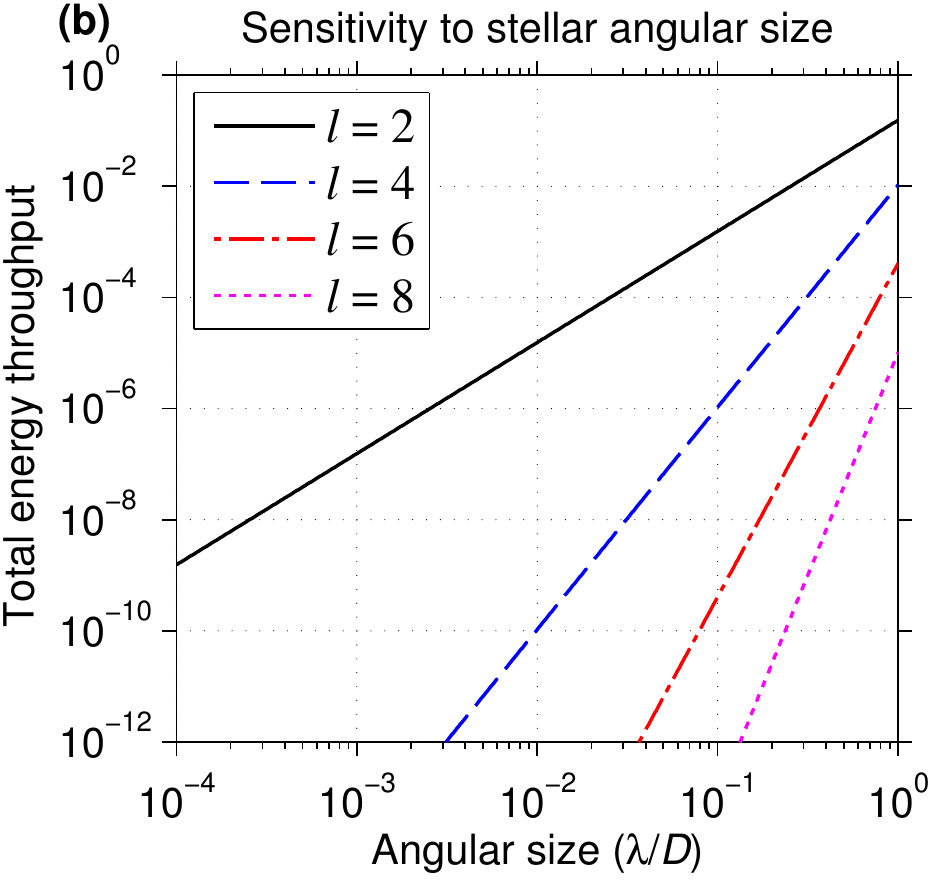}
    \includegraphics[trim={0.6cm 0 0 0},clip,height=0.322\linewidth]{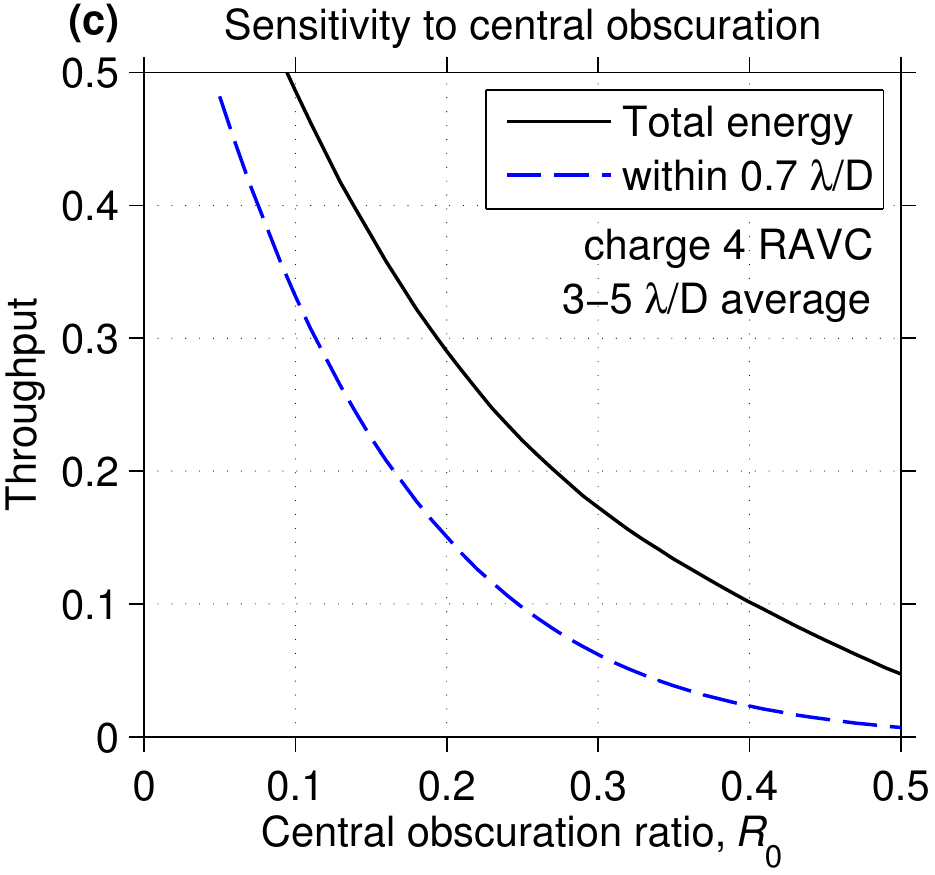}
    \caption{Sensitivity of the vortex coronagraph to (a) tip-tilt, (b) stellar angular size, and (c) central obscuration size. (a)-(b) The total energy throughput for an unobstructed circular pupil. (c) Throughput, averaged over source positions 3-5 $\lambda/D$, of a charge 4 ring-apodized vortex coronagraph (RAVC) for a simple annular aperture with central obscuration ratio $R_0$. Both the total energy and the energy within 0.7 $\lambda/D$ of the source position are shown.}
    \label{fig:baselinethpt}
\end{figure}

The value of the charge $l$ controls the off-axis throughput as well as sensitivity to tip-tilt, jitter, and stellar angular size (see Fig. \ref{fig:baselinethpt}a-b) \cite{Mawet2010b}. Although $l=2$ provides the best throughput for planets at small angular separations from their host stars, the performance will be heavily degraded on next-generation ground- and space-based telescopes owing to low-order aberrations and the partial resolution of the star. To hedge against this, we choose to use a minimum charge of $l=4$ in our designs. $l>4$ may be used in the future to reduce the sensitivity to such aberrations as needed at the cost of off-axis throughput.

The maximum throughput of an RAVC has a strong dependence on the size of the central obscuration. Fig.~\ref{fig:baselinethpt}c shows the throughput, averaged over angular separations 3-5~$\lambda/D$, as a function of $R_0$ in the $l=4$ case. Future segmented aperture telescopes, including the Thirty Meter Telescope (TMT) and potential space missions, will have $R_0$ values $\sim$14\%. The throughput loss with increasing $R_0$ is due to both the apodizer transmittance $t$ and the shape of the Lyot stop. 

We emphasize that the throughput values shown in Fig.~\ref{fig:baselinethpt}c are strictly for the RAVC and may be improved with alternate apodizer functions\cite{Carlotti2014} (Fogarty et al. 2016, in prep.) and/or modifications to the complex pupil field using low-loss techniques \cite{Pueyo2013,Mazoyer2015}. However, for the new coronagraphs presented below, the RAVC architecture serves as the initial condition of our apodizer optimization algorithm and therefore the throughput generally follows the trend shown. 

\subsection{Optimization method}
\label{subsec:method}
For arbitrary apertures, we use an iterative numerical optimization method to determine the optimal gray-scale apodizer to achieve a dark hole in the on-axis starlight at the final image plane. The problem is written in linear algebraic form as 
\begin{equation}
\underset{w}\min\left(||QCw||^2 + b||w - P A||^2\right),
\end{equation}
where $Q$ is a matrix that represents the dark hole region in the image plane, $C$ is the coronagraph operator that propagates the field from the apodizer plane to the final image plane, $P$ is the original telescope pupil function, $A$ is the current apodizer, and $w$ is the so-called auxiliary field in the apodizer plane which strikes a balance (as regulated by the weight $b$) between the field needed to generate a zero-valued dark hole and the physical field in the apodizer plane. Assuming the pupil and focal planes are related by Fourier transform propagation operators $F$, the coronagraph operator may be written $C=F \Theta F^{-1} \Omega F$, where $\Omega$ and $\Theta$ represent the focal plane mask and Lyot stop transmittance, respectively. The solution to the minimization problem is 
\begin{equation}
w=(b I + C^\dagger Q C)^{-1} b P A.
\end{equation}
To reduce computation time, the dimensionality of the inverted (square) matrix is reduced from the number of samples in the pupil plane to the number of samples in the dark hole region, by use of the Woodbury matrix identity:
\begin{equation}
w = \left[ I - C^\dagger Q(b Q + Q C C^\dagger Q)^{-1}Q C\right] P A.
\end{equation}
Since $w$ may be a complex function with infinite support, the physical apodizer is taken to be $A = |w|$ and $A$ is thresholded such that samples where $A>1$ are set to one and non-zero values outside of the original telescope pupil $P$ support are set to zero. A new auxiliary field is calculated based on the updated pupil field, and the process is repeated. The matrix $Q C C^\dagger Q$ is calculated once for each choice of focal plane mask, Lyot stop, and dark hole region as follows:
\begin{equation}
C C^\dagger = (F \Theta F^{-1} \Omega F) (F^{-1} \Omega^\dagger F \Theta^\dagger F^{-1}),
\end{equation}
\begin{equation}
C C^\dagger = F \Theta F^{-1}  |\Omega|^2 F \Theta^\dagger F^{-1}.
\end{equation}
Since the focal plane mask has phase-only transmittance (i.e. $|\Omega|^2=I$), this matrix simplifies to
\begin{equation}
C C^\dagger = F |\Theta|^2 F^{-1}
\end{equation}
and only depends on the squared modulus of Lyot stop function. 

We note that this algorithm may lead to very low throughput for poor choices of focal plane masks, Lyot stops, and apodizer initial conditions. The RAVC is a suitable initial design that leads to high throughput, even at at relatively small angular separations (potentially $<3~\lambda/D$). Another benefit of the RAVC, is that the numerical apodizer solution applies to all wavelengths.

In general, the starlight suppression in the dark hole region $Q$ is achieved at the cost of off-axis throughput. Assuming a good initial condition is chosen, the apodizer achieves a dark hole in the on-axis starlight without major losses. Finding the optimal combination of throughput and starlight suppression provided by the coronagraph masks requires performance metrics that relate these quantities to the sensitivity of the instrument to faint point sources in the presence of noise. 

\subsection{Optimization metrics: minimizing estimated integration time} 

The proposed optimization metrics are based on estimates of the signal-to-noise ratio (SNR) achieved in a given integration time for a typical planet, given by
\begin{equation}
\mathrm{SNR} = \frac{\epsilon \Delta t \Delta\lambda \eta_c\Phi_p}{\sqrt{\sigma^2_\mathrm{phot}+\sigma^2_\mathrm{det}+\sigma^2_\mathrm{spk}}},
\end{equation}
where the numerator represents the number of photo-electrons generated by planet light, $\epsilon = T q A_\mathrm{tel}$, $T$ is the telescope transmission, $q$ is the quantum efficiency, $A_\mathrm{tel}$ is collecting area of the telescope, $\Delta t$ is the effective integration time, $\Delta\lambda$ is the spectral bandwidth, $\eta_c$ is the coronagraph throughput, and $\Phi_p$ is the photon flux from the planet at the telescope aperture (photons per unit area per unit time per unit wavelength). 

The throughput of the coronagraph $\eta_c$ and the designed starlight suppression factor $s$ are defined as the fraction of planet and stellar energy that is incident on a single resolution element (a circle with assumed radius of 0.7 $\lambda/D$) for an ideal system without optical aberrations or atmospheric turbulence. Similarly, we define $\eta_0$ as the throughput without the coronagraphic masks in the system. 

The photon noise term may be expressed as $\sigma^2_\mathrm{phot} = \epsilon  \Delta t \Delta\lambda \hat{\Phi}$, where $\hat{\Phi} = \eta_c\Phi_p + s\Phi_\mathrm{star} + \Phi_b + \hat{\Phi}_\mathrm{spk}$ is the total flux within the resolution element, and $\Phi_\mathrm{star}$, $\Phi_b$, and $\hat{\Phi}_\mathrm{spk}$ are the photon fluxes owing to the stellar diffraction, background, and speckles, respectively. $\Phi_p$, $\Phi_\mathrm{star}$, and $\Phi_b$ are defined at the telescope aperture, whereas $\hat{\Phi}_\mathrm{spk}$ is defined in the image plane with respect to $\eta_0\Phi_\mathrm{star}$ and is treated separately from the diffracted starlight $s\Phi_\mathrm{star}$.

The detector noise is given by $\sigma^2_\mathrm{det} = i_d \Delta t + \sigma^2_\mathrm{read}$, where $i_d$ is the dark current and $\sigma^2_\mathrm{read}$ is the read noise averaged over many frames, which we approximate as $\sigma^2_\mathrm{read} \approx N_r^2 \epsilon \Delta t \Delta\lambda \hat{\Phi}_\mathrm{max}/W = \dot{R} \Delta t$, where $N_r$ is the read-out noise for each frame, $\hat{\Phi}_\mathrm{max}$ is the maximum photon flux in the image plane, and $W$ is the full well depth of the detector. 

The speckle flux is split into contributions owing to dynamic and quasi-static aberrations: $\hat{\Phi}_\mathrm{spk}=\hat{\Phi}_\mathrm{spk,dyn}+\hat{\Phi}_\mathrm{spk,qs}$. The former corresponds to residual wavefront distortions owing to the atmosphere with an average lifetime $\tau_\mathrm{dyn} = D/v$, where $D$ is outer diameter of the telescope aperture and $v$ is the wind speed. The latter is the slowly-varying aberration term owing to thermal and mechanical distortions in the system, which generate quasi-static speckles with lifetimes, $\tau_\mathrm{qs}$, on the order of hours. The effective speckle noise, excluding photon noise contributions, is approximated by
\begin{equation}
\sigma^2_\mathrm{spk} = \Delta t (\epsilon \Delta\lambda)^2 \tau_\mathrm{dyn} \mathrm{var}(\hat{\Phi}_\mathrm{spk}),
\end{equation} 
where
\begin{equation}
\tau_\mathrm{dyn}\mathrm{var}(\hat{\Phi}_\mathrm{spk})=\tau_\mathrm{dyn}\hat{\Phi}_\mathrm{spk,dyn}^2 + \tau_\mathrm{qs} \hat{\Phi}_\mathrm{spk,qs}^2 + 2 s \Phi_\mathrm{star} \left(\tau_\mathrm{dyn}\hat{\Phi}_\mathrm{spk,dyn} + \tau_\mathrm{qs}\hat{\Phi}_\mathrm{spk,qs}\right) + 2 \tau_\mathrm{dyn} \hat{\Phi}_\mathrm{spk,dyn} \hat{\Phi}_\mathrm{spk,qs},
\end{equation}
and $\mathrm{var}(\hat{\Phi}_\mathrm{spk})$ denotes the speckle flux variance, derived in Soummer et al. (2007) \cite{Soummer2007}. For a space telescope, the contribution of quickly-varying speckles is negligible, and the speckle noise term reduces to
\begin{equation}
\sigma^2_\mathrm{spk} = \Delta t (\epsilon \Delta\lambda)^2 \left( \tau_\mathrm{qs} \hat{\Phi}_\mathrm{spk,qs}^2 + 2 \tau_\mathrm{qs} s \Phi_\mathrm{star} \hat{\Phi}_\mathrm{spk,qs}\right).
\end{equation}

Recently developed methods to mitigate the speckle noise level via on-sky speckle nulling \cite{Martinache2014}, angular differential imaging \cite{Marois2006}, and sophisticated post-processing algorithms \cite{Soummer2012} significantly reduce the $\sigma^2_\mathrm{spk}$ term. We approximate these gains by using an effective quasi-static speckle flux level $g_\mathrm{sn} g_\mathrm{pp}\hat{\Phi}_\mathrm{spk,qs}$ in our calculations, where $g_\mathrm{sn}$ is the gain achieved through speckle nulling and $g_\mathrm{pp}$ is the gain achieved through speckle estimation and removal in post-processing. 

\begin{figure}[t]
    \centering
    \includegraphics[trim={0 0 1.6cm 0},clip,height=0.345\linewidth]{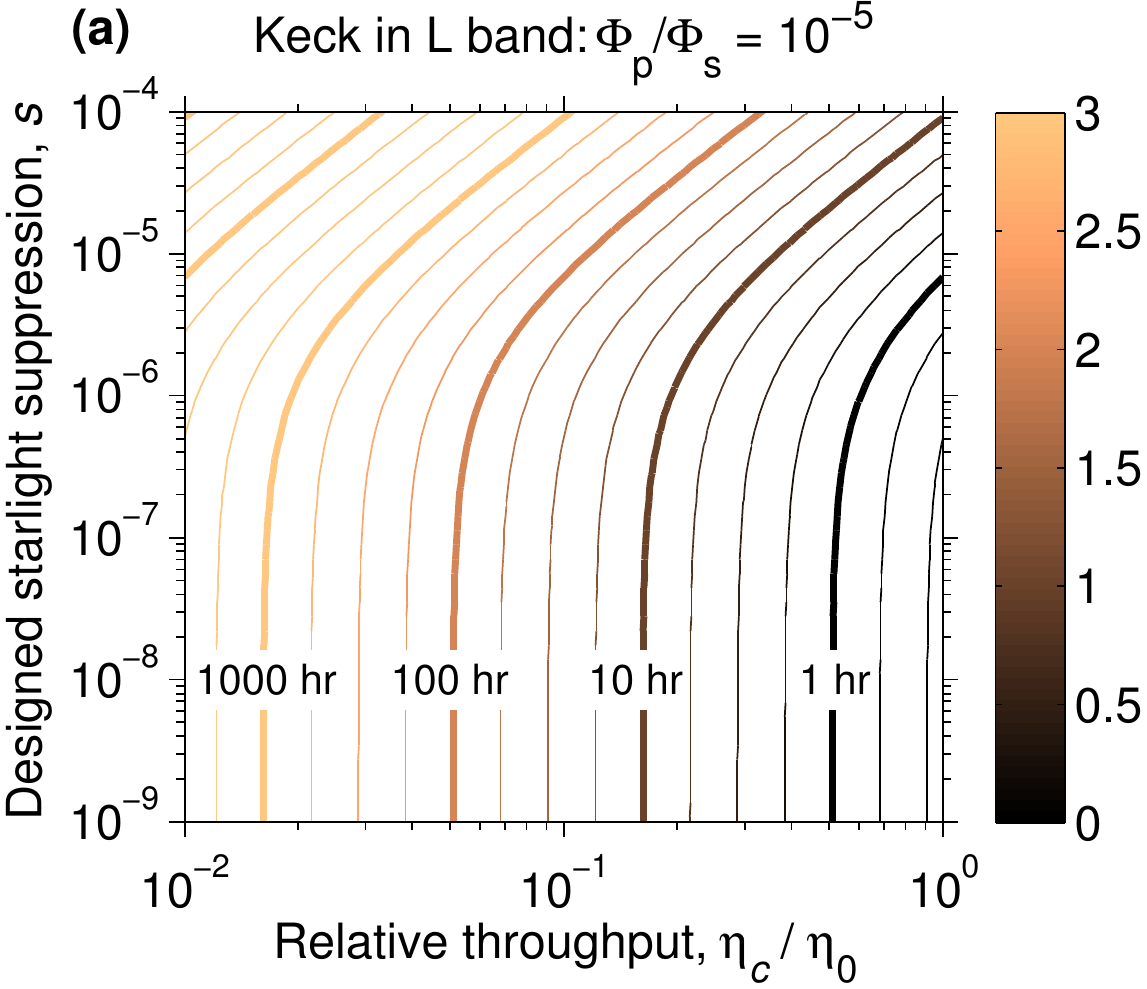}\hfill
    \includegraphics[trim={0.7cm 0 1.6cm 0},clip,height=0.345\linewidth]{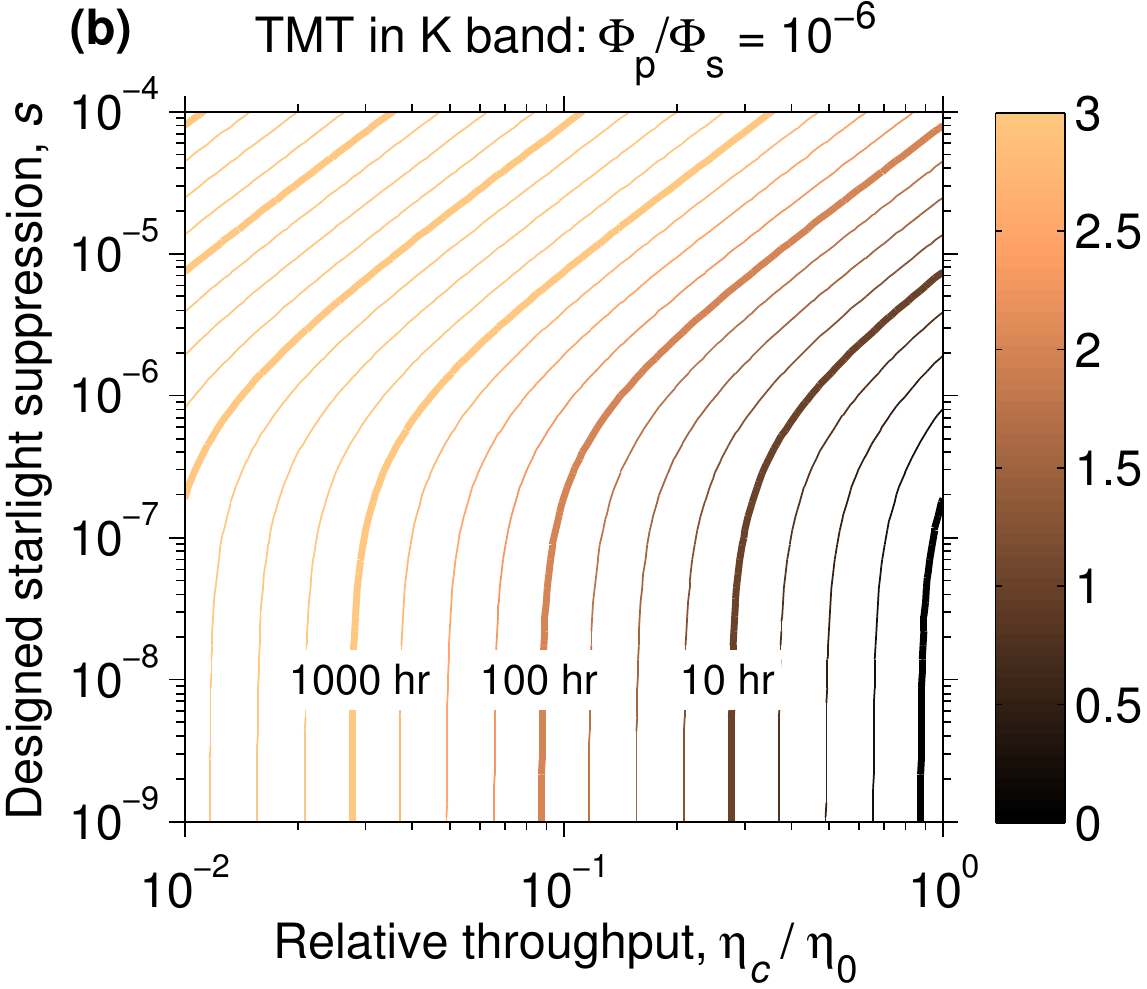}\hfill
    \includegraphics[trim={0.7cm 0 1.25cm 0},clip,height=0.345\linewidth]{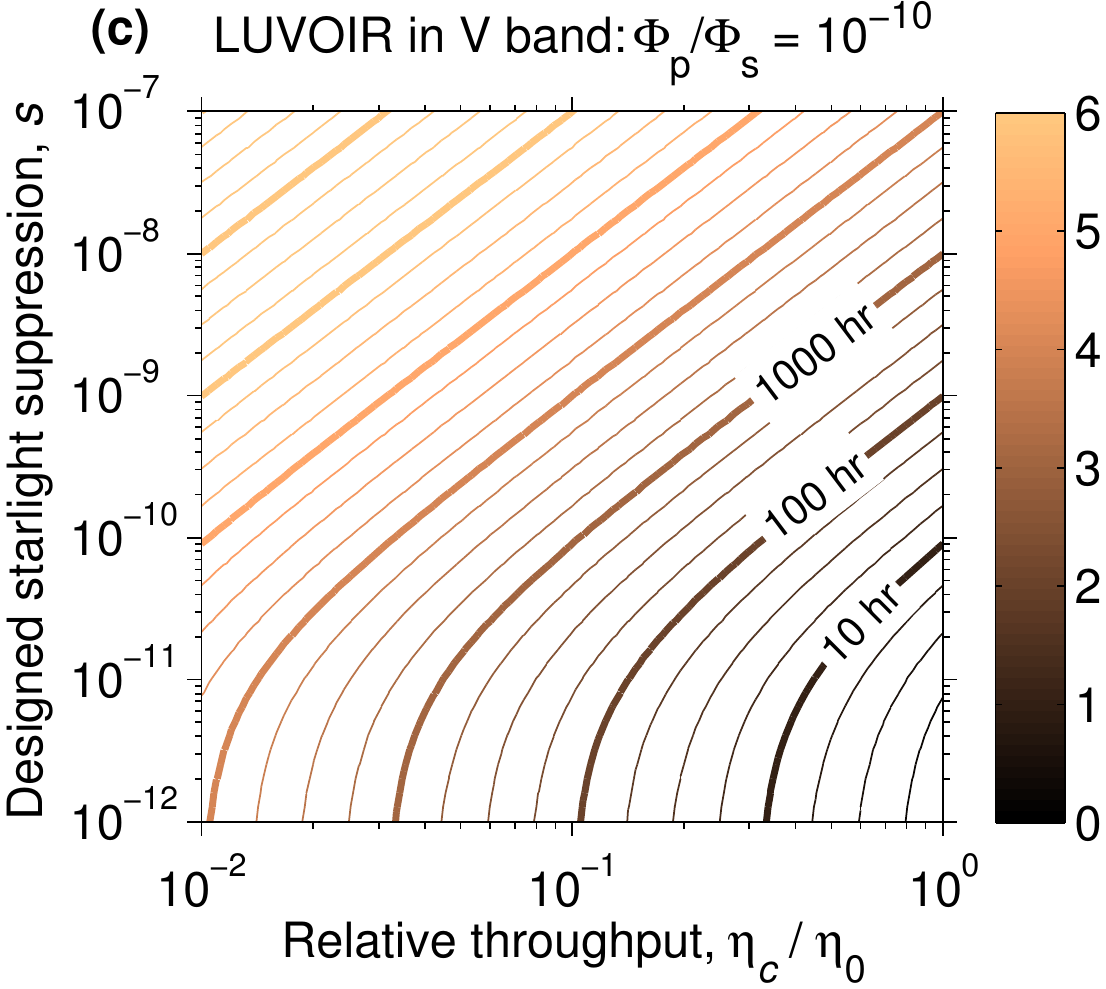}
    \caption{Estimates of integration time $\log_{10}(\Delta t)$ needed for a 5$\sigma$ detection with designed starlight suppression factor $s$ and throughput $\eta_c$ assuming planet:star contrast (a) $\Phi_p/\Phi_\mathrm{star}=10^{-5}$ with Keck in L band, (b) $\Phi_p/\Phi_\mathrm{star}=10^{-6}$ with TMT in K band, and (c) $\Phi_p/\Phi_\mathrm{star}=10^{-10}$ with a future LUVOIR space telescope in V band. $\eta_c$ and $s$ are solely properties of the coronagraph design and refer to the performance under ideal conditions. The effect of $s$ on speckle noise is described by Eqn. \ref{eqn:dt_spk}. The assumptions used for these calculations may be found in the appendix.}
    \label{fig:exp_time}
\end{figure}
\newpage
Since $\sigma^2_\mathrm{phot}$, $\sigma^2_\mathrm{det}$, and $\sigma^2_\mathrm{spk}$ increase linearly with $\Delta t$ under these assumptions, we can solve for the integration time needed to reach an SNR threshold $\Gamma$:
\begin{equation}
\Delta t = \Gamma^2\left(\Delta t_\mathrm{phot} + \Delta t_\mathrm{det} + \Delta t_\mathrm{spk}\right),
\end{equation}
where $\Delta t_\mathrm{phot}$, $\Delta t_\mathrm{det}$, and $\Delta t_\mathrm{spk}$ are the scale times to overcome each noise term:
\begin{equation}
\Delta t_\mathrm{phot} = \frac{1}{(\eta_c \Phi_p)^2}\frac{\hat{\Phi}}{\epsilon \Delta\lambda},
\end{equation}
\begin{equation}
\Delta t_\mathrm{det} = \frac{1}{(\eta_c \Phi_p)^2}\frac{\dot{R} + i_d }{(\epsilon \Delta\lambda)^2},
\end{equation}
\begin{equation}
	\Delta t_\mathrm{spk}=\frac{1}{(\eta_c \Phi_p)^2}\left\{ \begin{array}{ll}
   \tau_\mathrm{dyn} \mathrm{var}(\hat{\Phi}_\mathrm{spk}) & \text{for ground-based telescopes,}  \\
   \tau_\mathrm{qs} \hat{\Phi}_\mathrm{spk,qs}^2 + 2 \tau_\mathrm{qs} s \Phi_\mathrm{star} \hat{\Phi}_\mathrm{spk,qs} & \text{for space telescopes.}  \\
\end{array} \right.
\label{eqn:dt_spk}
\end{equation}

In the calculations shown in Fig. \ref{fig:exp_time}, it can be seen designing the coronagraph masks to have a smaller value of $s$ leads to shorter integration times until the dominant speckle level is reached. The integration time scales as 
\begin{equation}
	\Delta t \propto \left\{ \begin{array}{ll}
   s/\eta_c^2 & s>\hat{\Phi}_\mathrm{spk,0}/ (2 \eta_0 \Phi_\mathrm{star})  \\
   1/\eta_c^2 & s<\hat{\Phi}_\mathrm{spk,0}/ (2 \eta_0 \Phi_\mathrm{star}) \\
\end{array} \right.,
\end{equation}
where $\hat{\Phi}_\mathrm{spk,0}$ is the dominant speckle flux term: $\hat{\Phi}_\mathrm{spk,0} \sim \hat{\Phi}_\mathrm{spk,dyn}$ for ground-based and $\hat{\Phi}_\mathrm{spk,0} \sim \hat{\Phi}_\mathrm{spk,qs}$ for space-based applications. This transition occurs at $s \sim 10^{-6}-10^{-5}$ for ground-based telescopes and $s \sim 10^{-11}-10^{-10}$ in space. The integration time depends strongly on $\eta_c$ in both regimes and therefore the throughput is the single most important coronagraph design parameter for minimizing the integration time.

\section{Apodized vortex coronagraphs for segmented apertures} \label{sec:solutions} 

Using the method outlined in section \ref{subsec:method}, we optimized apodizing pupil masks for a charge 4 vortex coronagraph on the Thirty Meter Telescope (TMT) and potential future space telescopes \cite{SCDA}. 

\begin{figure}[t]
    \centering
    \includegraphics[trim={0 0 0 0},clip,height=0.28\linewidth]{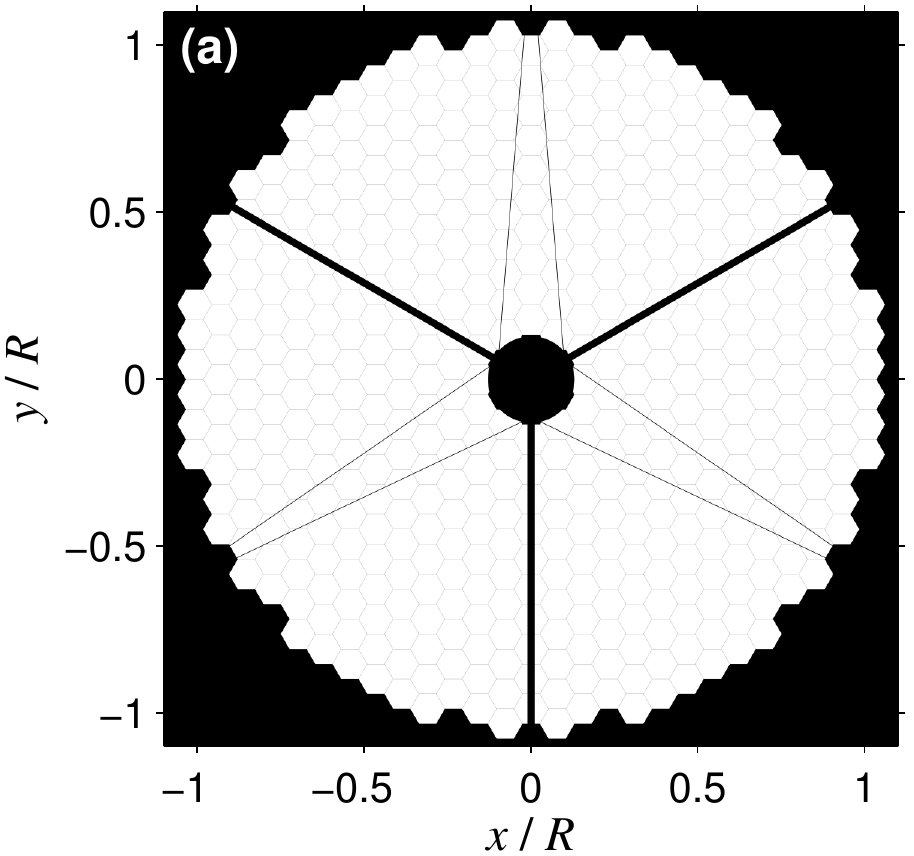}\hfill
    \includegraphics[trim={0 0 0 0},clip,height=0.28\linewidth]{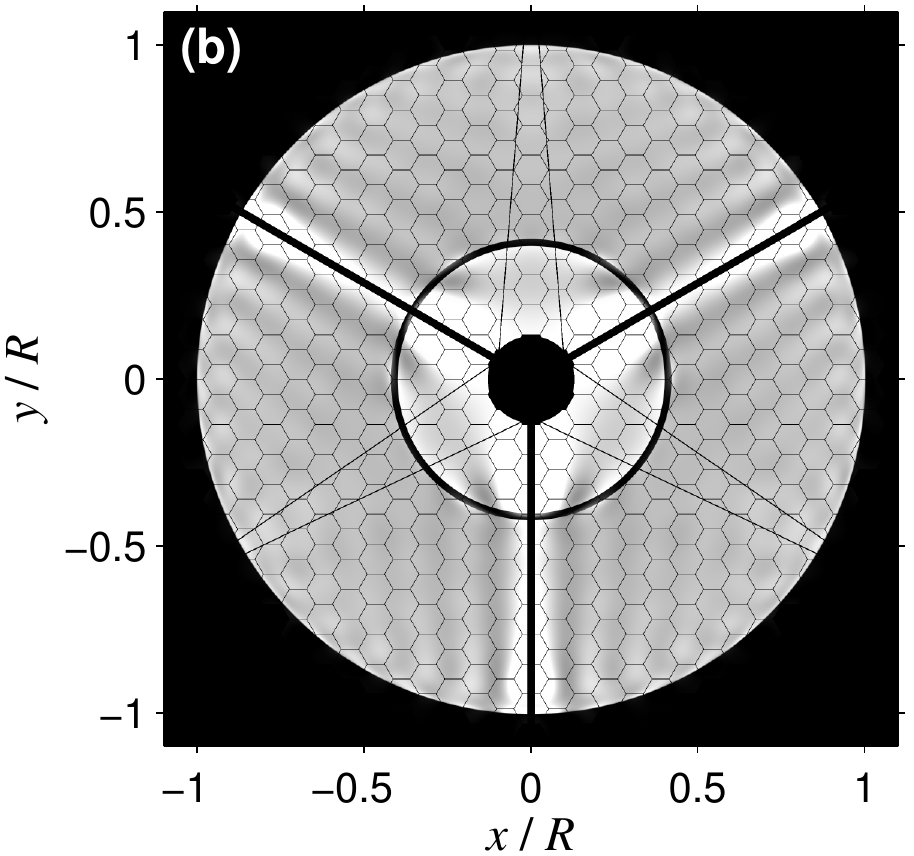}\hfill
    \includegraphics[trim={0 0 0 0},clip,height=0.28\linewidth]{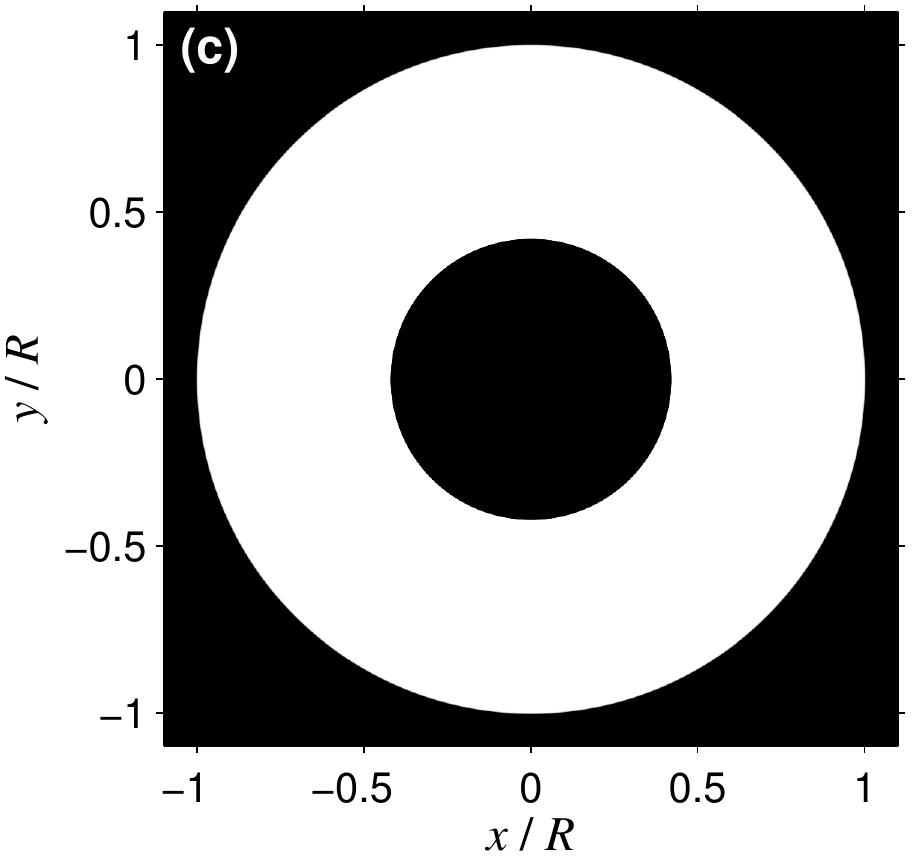}\hspace{1mm}
    \includegraphics[trim={0 -8mm -5mm 0 },clip,scale=0.65]{colorbarsforfig3.pdf}\\
    \includegraphics[height=0.28\linewidth]{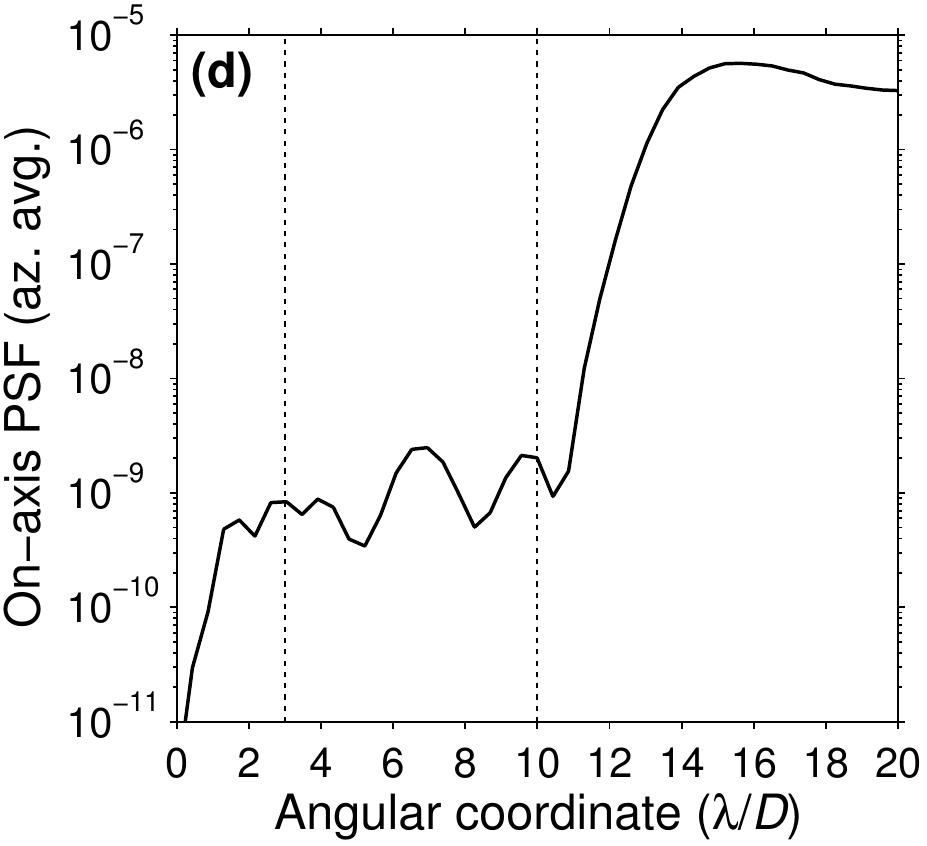}\hfill
    \includegraphics[height=0.275\linewidth]{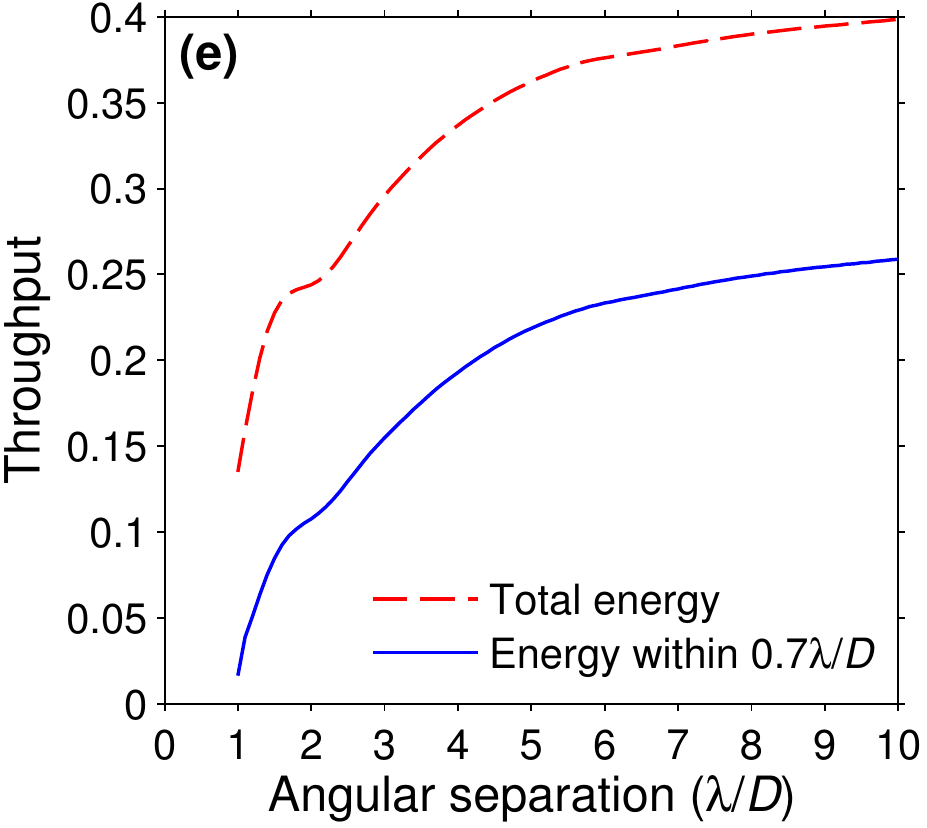}\hfill
    \includegraphics[height=0.28\linewidth]{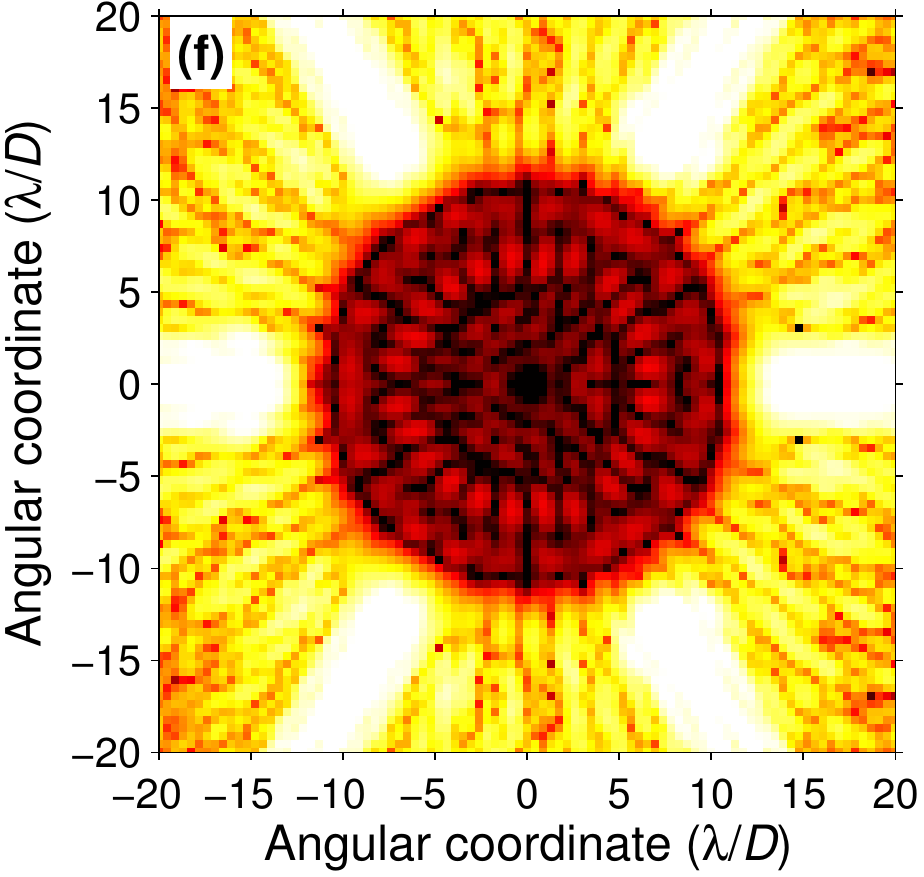}
    \includegraphics[trim={0 -8mm 0 0},clip,scale=0.65]{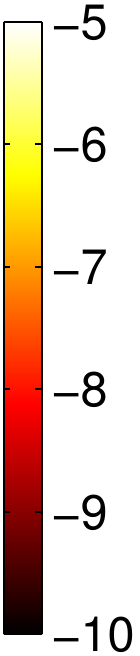}\\
    \caption{Apodizer for Thirty Meter Telescope (TMT) aperture. (a)~Pupil amplitude $P$ prior to the apodizer. (b)~Gray-scale apodizer mask $A$ in pupil. (c)~Lyot stop $\Omega$. (d)~Azimuthal average of the on-axis PSF. (e)~Throughput of the coronagraph masks, normalized the to telescope throughput. (f)~On-axis PSF (log irradiance) for an ideal wavefront, normalized to the peak of the telescope PSF, without optical aberrations.}
    \label{fig:TMT}
\end{figure}

\subsection{Apodizer for the Thirty Meter Telescope (TMT)}

The goal of ground-based coronagraphic instruments is to minimize integration time by reducing the residual starlight factor $s$ below the dynamic speckle noise level, while maintaining as much throughput as possible. For TMT, we assume $\hat{\Phi}_\mathrm{spk,dyn}=10^{-5}\eta_0\Phi_\mathrm{star}$ and therefore pinned speckles are sufficiently mitigated if $s \ll 5\times10^{-6}$. We design our masks to achieve an ideal value of $s = 10^{-9}$, which leaves a considerable margin for additional slowly-varying speckles caused by manufacturing errors and unforeseen aberrations in the optical system. We also find that the resulting throughput is relatively insensitive to the design value of $s$ when an RAVC is used as the initial condition ($\sim\!6\%$ loss with respect to the initial RAVC throughput).

An apodizer solution for TMT and its corresponding performance is shown in Fig. \ref{fig:TMT}. The coronagraph is made up of three masks: a gray-scale apodizer $A$ in the pupil (see Fig. \ref{fig:TMT}a-b), a charge 4 vortex phase mask in the focal plane $\Omega$, and an annular Lyot stop $\Theta$ (see Fig. \ref{fig:TMT}c). The azimuthal average of the on-axis PSF, shown in Fig. \ref{fig:TMT}d, is at the $\sim\!10^{-9}$ level throughout the dark hole when normalized to the telescope PSF, and therefore $s \approx 10^{-9}$ within an annulus from 3 to 10~$\lambda/D$. The throughput (see Fig. \ref{fig:TMT}e) increases from small to large angular separations across the dark hole region. The energy within the planet PSF core is greater than 15\% for angular separations $>3~\lambda/D$. Although the throughput is calculated for point sources displaced along a single direction, the throughput should be approximately the same in all directions, and the starlight is uniformly suppressed throughout the dark hole (see Fig. \ref{fig:TMT}f). 

\begin{figure}[p]
    \centering
    \includegraphics[trim={0 0 0 0},clip,height=0.28\linewidth]{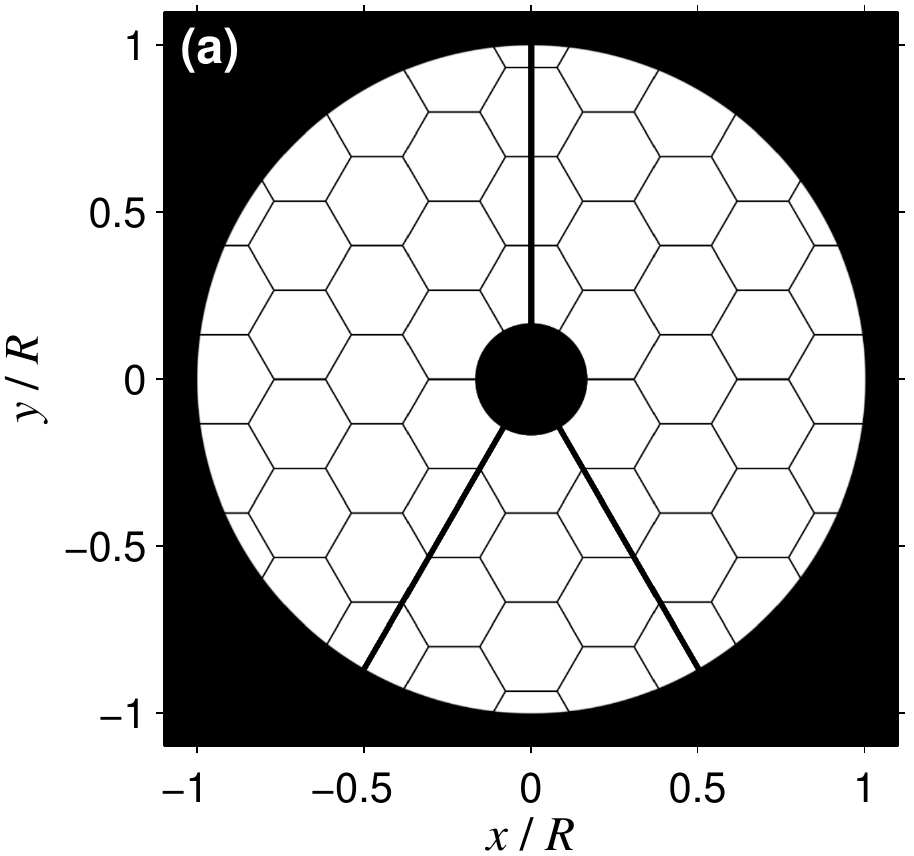}\hfill
    \includegraphics[trim={0 0 0 0},clip,height=0.28\linewidth]{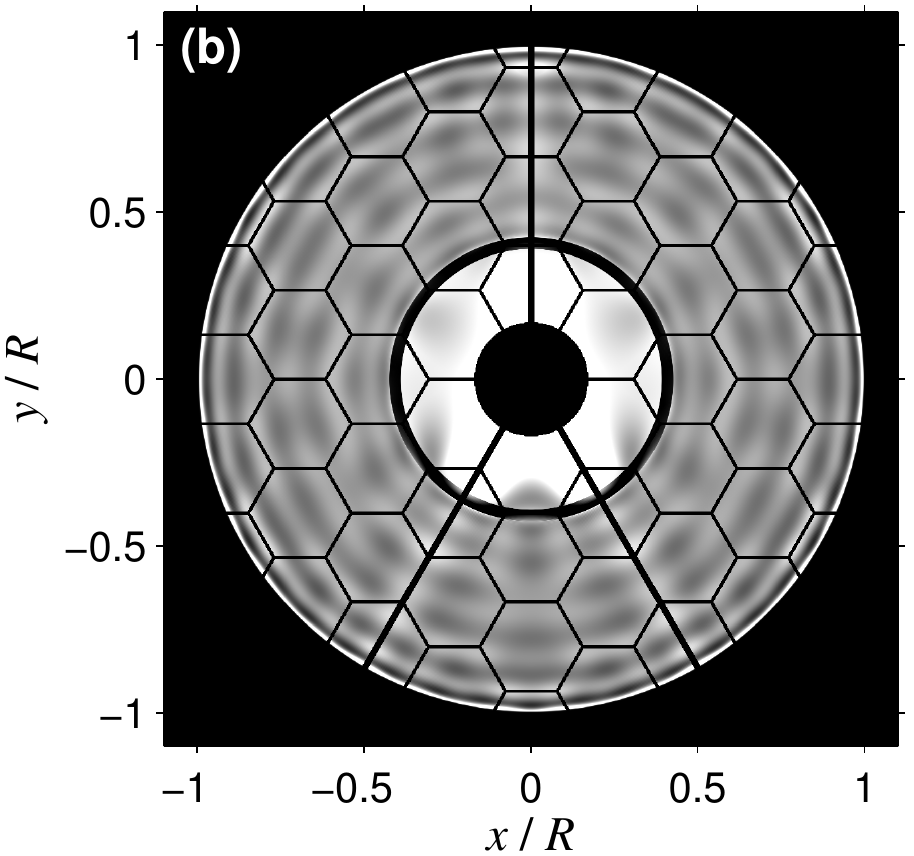}\hfill
    \includegraphics[trim={0 0 0 0},clip,height=0.28\linewidth]{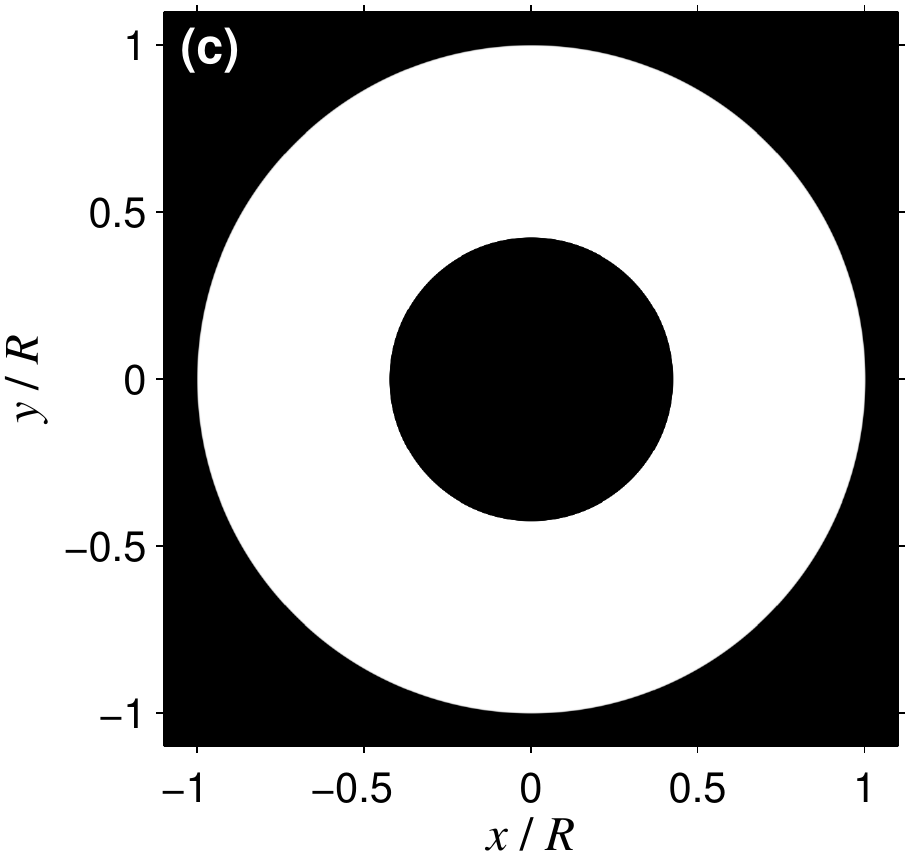}\hspace{1mm}
    \includegraphics[trim={0 -8mm -5mm 0 },clip,scale=0.65]{colorbarsforfig3.pdf}\\
    \includegraphics[height=0.28\linewidth]{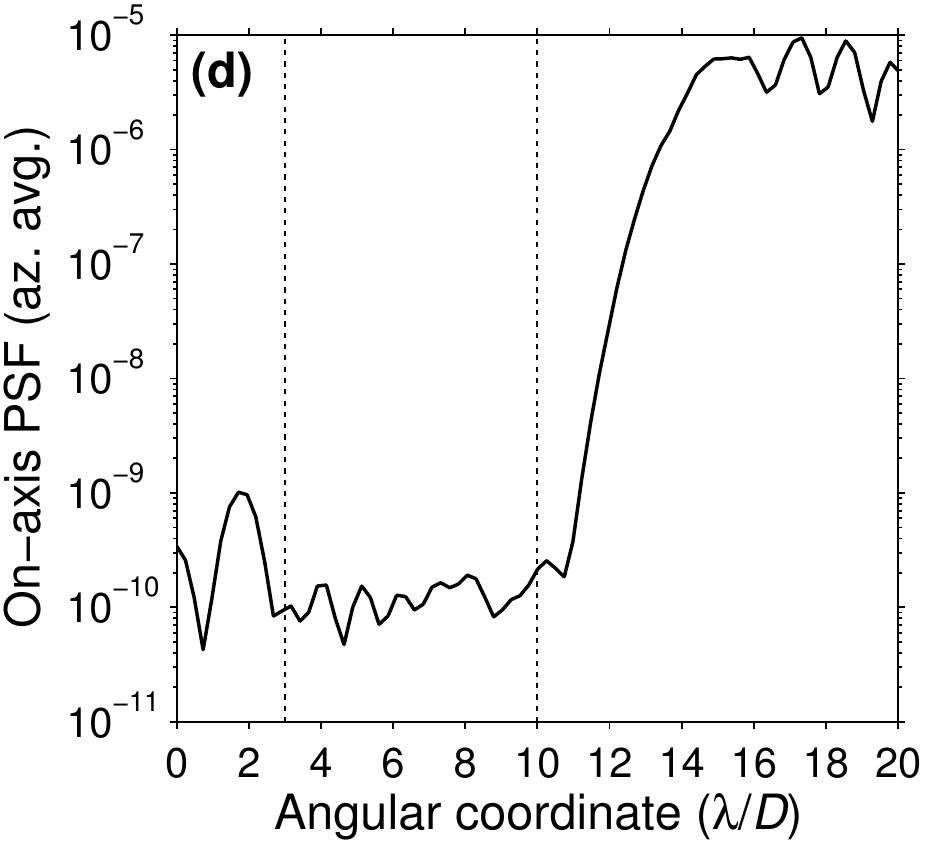}\hfill
    \includegraphics[height=0.275\linewidth]{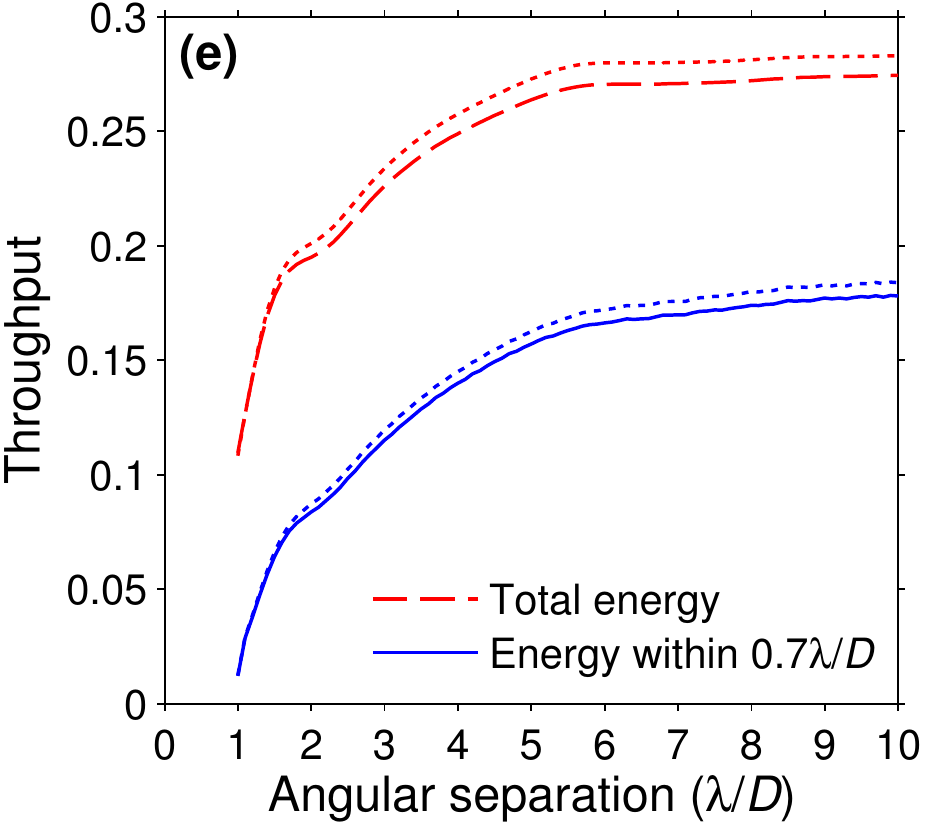}\hfill
    \includegraphics[height=0.28\linewidth]{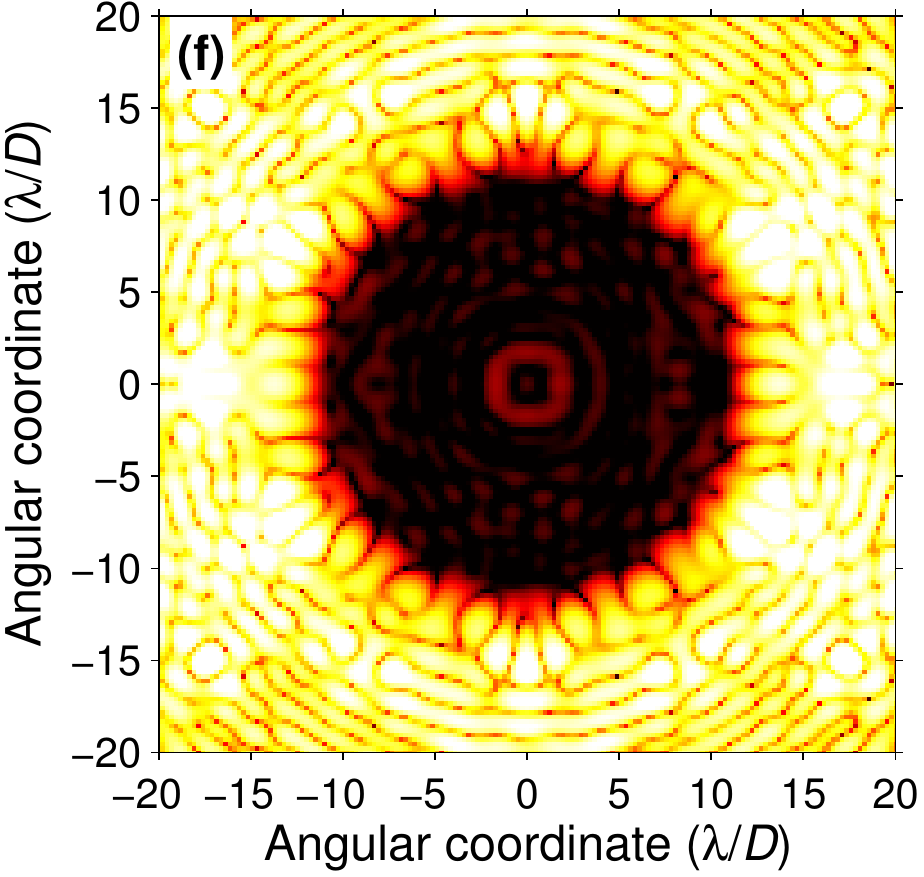}
    \includegraphics[trim={0 -8mm 0 0},clip,scale=0.65]{colorbarsforfig5.pdf}\\
    \caption{Same as Fig. \ref{fig:TMT}, but for a segmented aperture with four-rings of hexagonal mirrors and thick spiders. The dotted lines in (e) show the throughput achieved with the spiders removed.}
    \label{fig:clippedhex4_w_spiders}
    \vspace*{\floatsep}
    \includegraphics[trim={0 0 0 0},clip,height=0.28\linewidth]{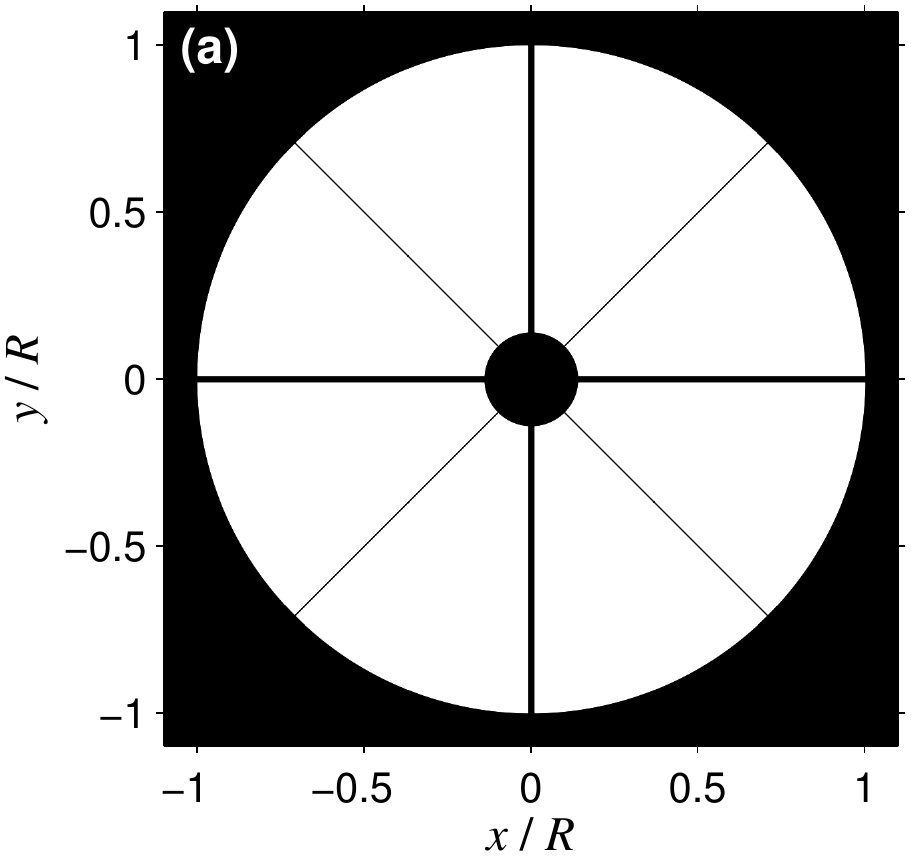}\hfill
    \includegraphics[trim={0 0 0 0},clip,height=0.28\linewidth]{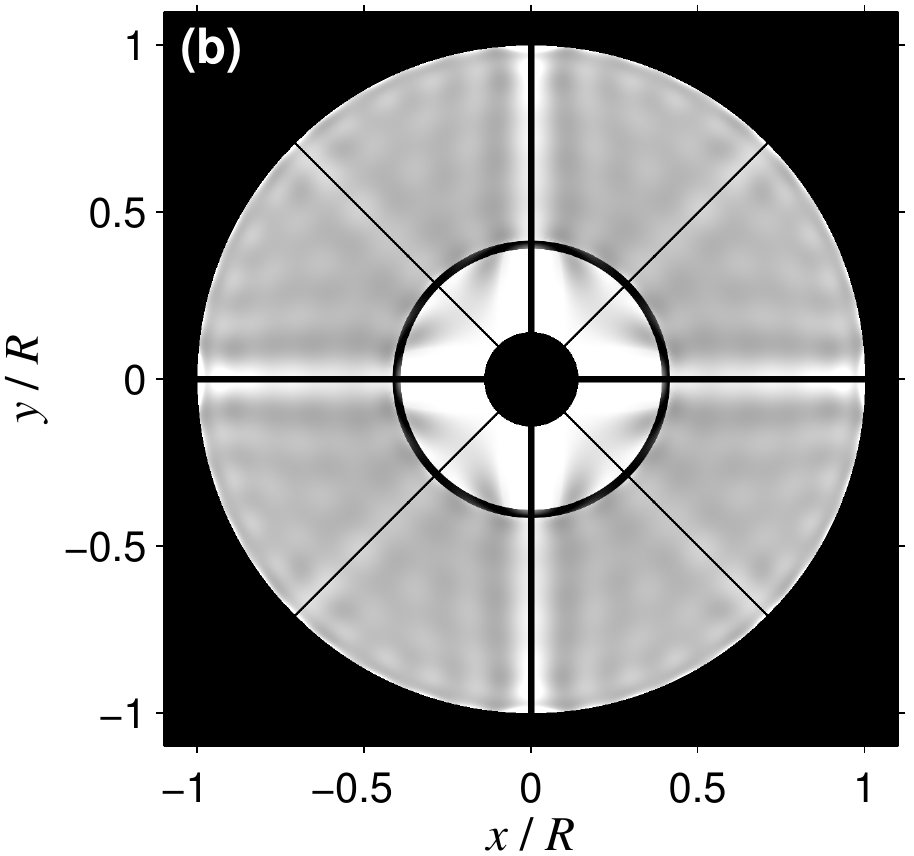}\hfill
    \includegraphics[trim={0 0 0 0},clip,height=0.28\linewidth]{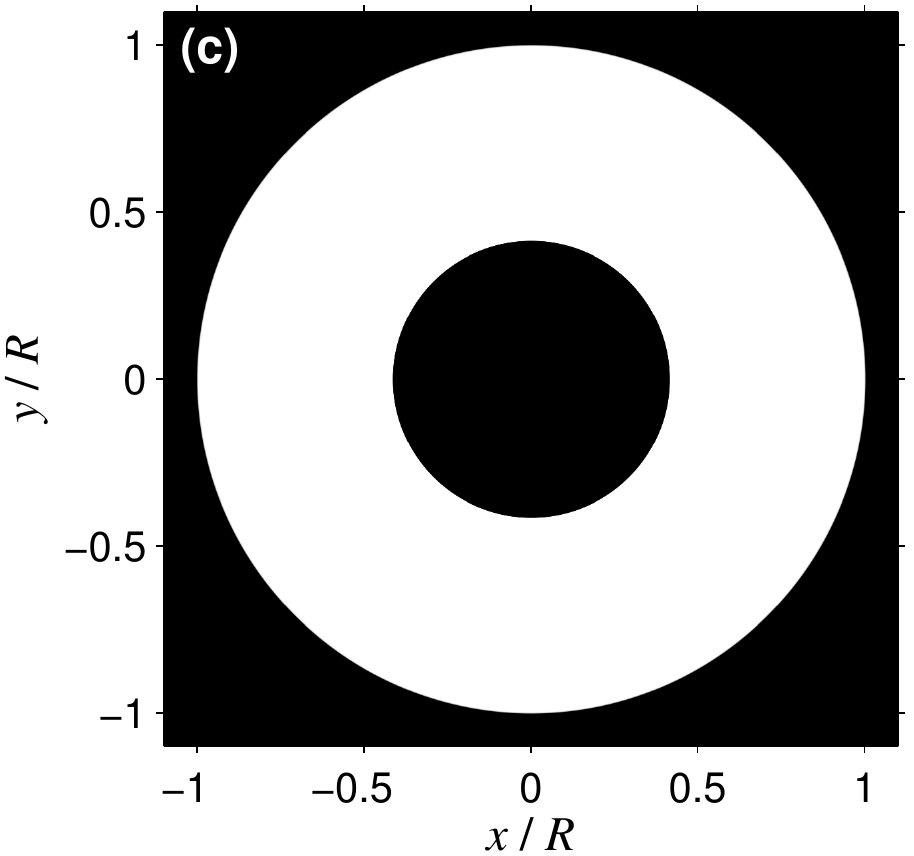}\hspace{1mm}
    \includegraphics[trim={0 -8mm -5mm 0 },clip,scale=0.65]{colorbarsforfig3.pdf}\\
    \includegraphics[height=0.28\linewidth]{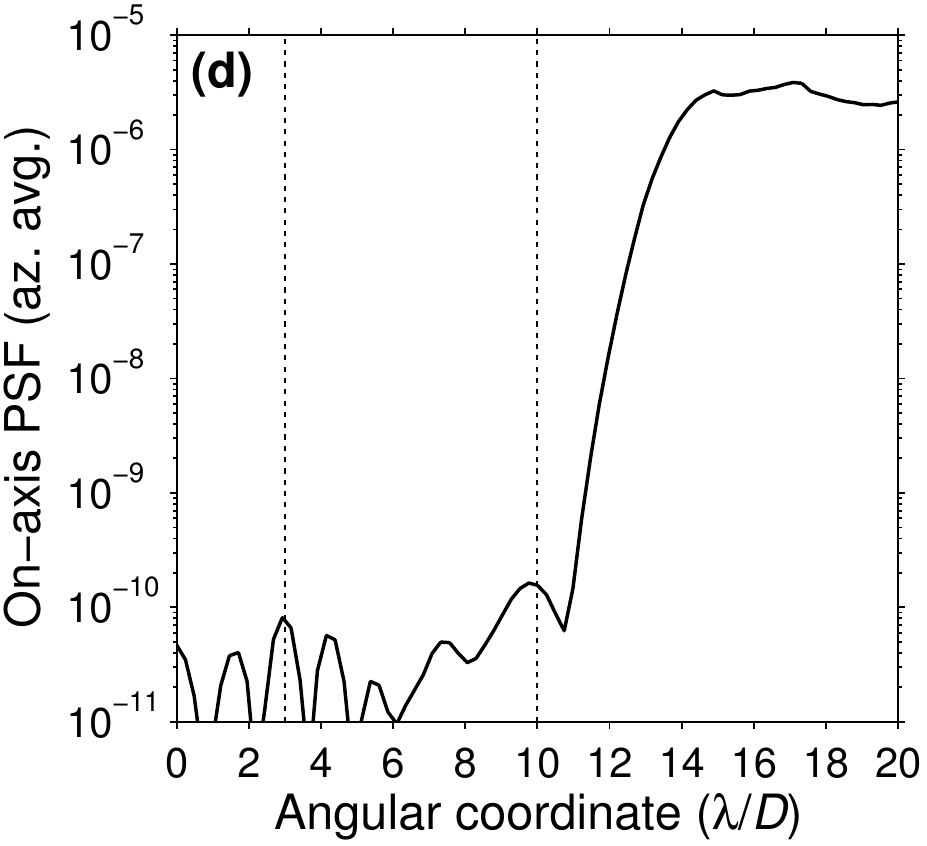}\hfill
    \includegraphics[height=0.275\linewidth]{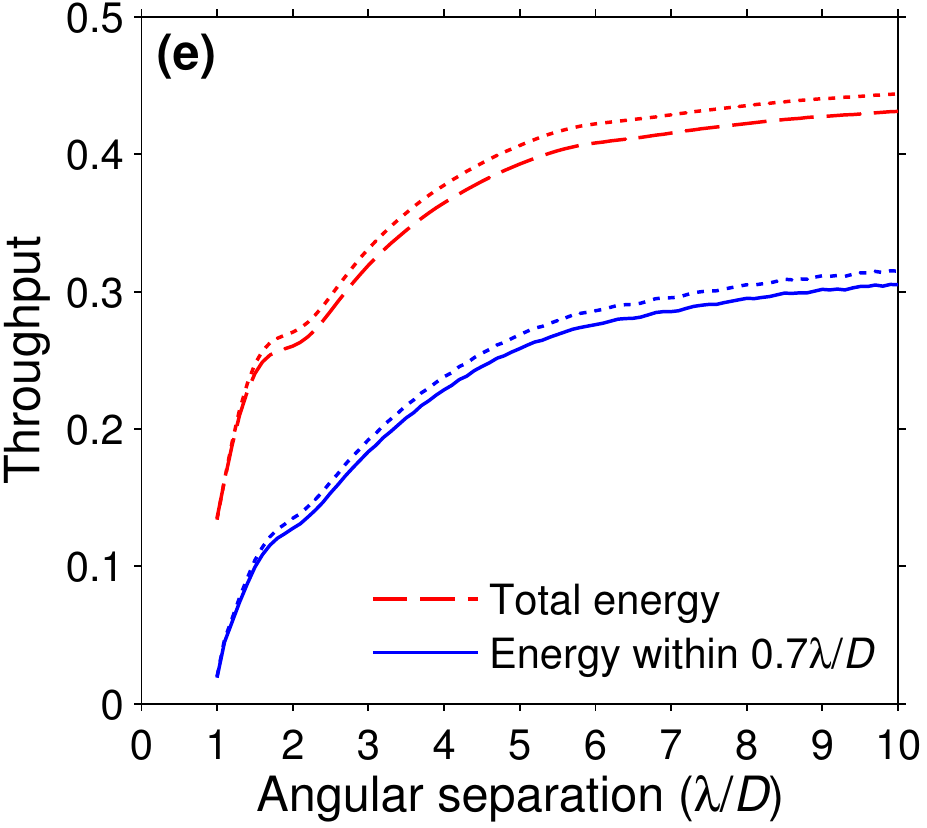}\hfill
    \includegraphics[height=0.28\linewidth]{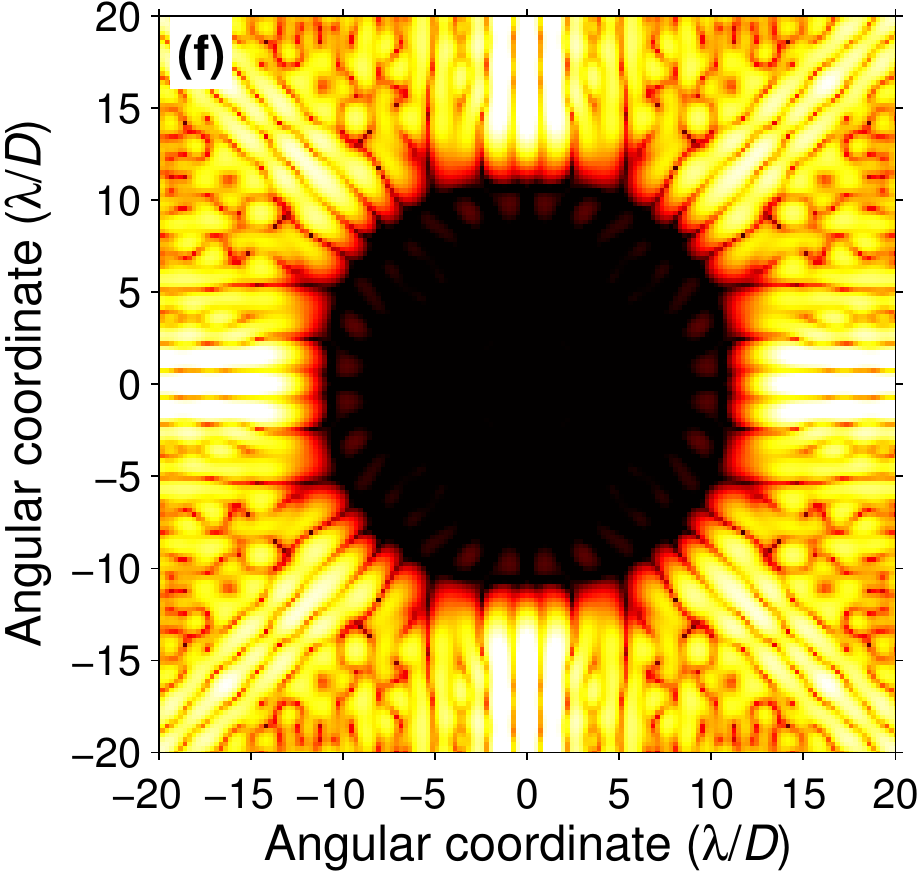}
    \includegraphics[trim={0 -8mm 0 0},clip,scale=0.65]{colorbarsforfig5.pdf}\\
    \caption{Same as Fig. \ref{fig:clippedhex4_w_spiders}, but for a segmented aperture with eight pie-wedge mirrors and thick spiders. The dotted lines in (e) show the throughput achieved with the spiders removed.}
    \label{fig:piewedge8_w_spiders}
\end{figure}

\subsection{Apodizers for future segmented-aperture space telescopes}

A major driver for investing in coronagraphic instruments on space telescopes is that the speckle level will be far lower without the contributions from the quickly varying atmosphere ($\hat{\Phi}_\mathrm{spk,dyn}=0$ and $\hat{\Phi}_\mathrm{spk,qs}=10^{-9}\eta_0\Phi_\mathrm{star}$). Therefore, the goal for space-based applications is to reduce $s$ as low as $10^{-10}$ at the highest possible throughput. Here, we consider two potential aperture types for future segmented space telescopes: a segmented primary with four-rings of hexagonal mirrors (see Fig. \ref{fig:clippedhex4_w_spiders}) and one with eight pie-wedge mirrors (see Fig. \ref{fig:piewedge8_w_spiders}). The spiders in each case are chosen to be co-aligned with discontinuities between segments.   

In the case of the hexagonally-segmented pupil (Fig.~\ref{fig:clippedhex4_w_spiders}), the optimized apodizer reduces the starlight to the $10^{-10}$ level while providing 12\% PSF core throughput at 3~$\lambda/D$. In the case of the pie-wedge aperture (Fig.~\ref{fig:piewedge8_w_spiders}), the PSF core throughput is 20\% at 3~$\lambda/D$. The gain in throughput is mostly attributed to the size of central obscuration, which is slightly smaller in the case of the pie-wedge aperture ($R_0=0.17$ versus $R_0=0.14$). 
We find that the discontinuities in the pupil have a much smaller effect on the throughput achieved. At $s\approx10^{-10}$, the throughput difference with and without spiders is only a couple of percent, as indicated by the dotted lines in Figs. \ref{fig:clippedhex4_w_spiders}e and \ref{fig:piewedge8_w_spiders}e.

All of the apodized vortex solutions shown are theoretically broadband; that is, the suppression and throughput shown are expected at all wavelengths provided the masks have constant complex transmittance over the passband. The practical broadband performance will only be limited by manufacturing constraints. 

\newpage
\section{CONCLUSIONS AND FUTURE OUTLOOK} \label{sec:concl} 

We have presented apodized pupil vortex coronagraphs designed for ground- and space-based telescopes with segmented apertures. in each case, the coronagraph masks are optimized such that the estimated integration time is minimized, in the presence of noise.

The design goals in terms of starlight suppression and throughput for ground- and space-based applications depend mostly strongly on the expected speckle noise characteristics. In case of TMT, the coronagraph masks suppress diffracted starlight to the $10^{-9}$ level, assuming no aberrations in the system, which is well below the expected speckle noise level. For space telescopes, we show solutions that push the diffraction down to the $10^{-10}$ level while maintaining sufficient throughput to significantly reduce integration time estimates. 

The throughput achieved is relatively insensitive to aperture discontinuities. In the case of the space telescopes, the throughput with and without thick spiders only differs by a couple of percent. However, the central obscuration size has a much larger effect. The best performance is expected for telescopes with relatively small secondary mirrors. 

Although the theoretical solutions are independent of wavelength, manufacturing errors will ultimately limit the performance of these coronagraph designs. Pathways to approach the ideal performance are available thanks to successful demonstrations of broadband vortex phase masks based on liquid crystal polymers \cite{Mawet2009} and sub-wavelength gratings \cite{Delacroix2013} as well as gray-scale ring apodizers for vortex coronagraphs \cite{Mawet2014}. Detailed simulations are underway to study the chromaticity of gray-scale apodizers fabricated using various methods. Outcomes of these studies will inform a second generation of the presented coronagraph designs that incorporate known material properties. 

The methods employed here may be readily generalized to include optimization of the deformable mirror shapes to achieve broadband starlight suppression, potentially at high throughput. A comprehensive exploration of apodizer solutions and designs that also make use two deformable mirrors will be the topic of an upcoming paper.

\newpage
\appendix

\section{Assumptions used in integration time calculations}

\begin{table}[h]
\caption{Notional instrument parameters of planet finding instruments for Keck, TMT, and LUVOIR telescopes.}
\label{tab:fonts}
\begin{center}       
\begin{tabular}{|l|l|l|l|l|} 
\hline
\rule[-1ex]{0pt}{3.5ex} Quantity & Symbol & Keck (L) & TMT (K) & LUVOIR (V)\\
\hline\hline
\rule[-1ex]{0pt}{3.5ex}  Aperture diameter (m) & $D$ & 10 & 30 & 12 \\
\hline
\rule[-1ex]{0pt}{3.5ex}  Central obscuration ratio & $R_0$ & 0.3 & 0.13 & 0.14 \\
\hline
\rule[-1ex]{0pt}{3.5ex}  Aperture collecting area (m$^2$) & $A_\mathrm{tel}$ & 38 & 535 & 84 \\
\hline
\rule[-1ex]{0pt}{3.5ex}  Central wavelength ($\mu$m) & $\lambda_0$ & 3.78 & 2.20 & 0.55\\
\hline
\rule[-1ex]{0pt}{3.5ex}  Bandwidth ($\mu$m) & $\Delta \lambda$ & 0.7 & 0.4 & 0.05\\
\hline
\rule[-1ex]{0pt}{3.5ex}  Telescope transmission & $T$ & 0.5 & 0.5 & 0.5  \\
\hline
\rule[-1ex]{0pt}{3.5ex}  Non-coronagraphic throughput  & $\eta_0$ & 0.6 & 0.6 & 0.6  \\
\hline
\rule[-1ex]{0pt}{3.5ex}  Quantum efficiency (e-/photon) & $q$ & 0.8 & 0.8 & 0.8 \\
\hline
\rule[-1ex]{0pt}{3.5ex}  Dark current (e-/pixel/sec) & $i_d$ & 0.1 & 0.01 & 0.005\\
\hline
\rule[-1ex]{0pt}{3.5ex}  Read noise (e-/pixel/sec) & $N_r$ & 15 & 5 & 1 \\
\hline 
\rule[-1ex]{0pt}{3.5ex}  Well depth (e-) & $W$ & 18,000 & 25,000 & 50,000 \\
\hline
\rule[-1ex]{0pt}{3.5ex}  Apparent magnitude of star & - & $L=4$ & $K=4$ & $V=4$\\
\hline
\rule[-1ex]{0pt}{3.5ex}  Stellar flux (photons/sec/m$^2$/$\mu$m) & $\Phi_\mathrm{star}$ & $2.50\times10^7$ & $1.14\times10^8$ & $2.41\times10^9$    \\
\hline
\rule[-1ex]{0pt}{3.5ex}  Planet flux (photons/sec/m$^2$/$\mu$m) & $\Phi_p$ & $10^{-5} \Phi_\mathrm{star}$ & $10^{-6} \Phi_\mathrm{star}$ & $10^{-10} \Phi_\mathrm{star}$ \\
\hline 
\rule[-1ex]{0pt}{3.5ex}  Background flux (photons/sec/m$^2$/$\mu$m) & $\Phi_b$ & 7390 & 0.11 & 0   \\
\hline 
\rule[-1ex]{0pt}{3.5ex}  Dynamic speckle flux (photons/sec/m$^2$/$\mu$m) & $\hat{\Phi}_\mathrm{speck,f}$ & $10^{-4} \eta_0\Phi_\mathrm{star}$ & $10^{-5} \eta_0\Phi_\mathrm{star}$ & -   \\
\hline 
\rule[-1ex]{0pt}{3.5ex}  Quasi-static speckle flux (photons/sec/m$^2$/$\mu$m) & $\hat{\Phi}_\mathrm{speck,s}$ & $10^{-5} \eta_0\Phi_\mathrm{star}$ & $10^{-6} \eta_0\Phi_\mathrm{star}$ & $10^{-9} \eta_0\Phi_\mathrm{star}$   \\
\hline 
\rule[-1ex]{0pt}{3.5ex} Wind speed (m/s) & $v$ & 10 & 10 & -   \\
\hline 
\rule[-1ex]{0pt}{3.5ex} Dynamic speckle lifetime (sec) & $\tau_\mathrm{f}$ & $D/v=1$ & $D/v=3$ & -   \\
\hline 
\rule[-1ex]{0pt}{3.5ex} Quasi-static speckle lifetime (sec) & $\tau_\mathrm{s}$ & 3600 & 3600 & 3600  \\
\hline 
\rule[-1ex]{0pt}{3.5ex}  Speckle nulling improvement factor & $g_\mathrm{sn}$ & $1/3$ & $1/3$ & -  \\
\hline 
\rule[-1ex]{0pt}{3.5ex}  Post-processing improvement factor & $g_\mathrm{pp}$ & $1/10$ & $1/10$ & $1/30$   \\
\hline 
\rule[-1ex]{0pt}{3.5ex}  Maximum flux (photons/sec/m$^2$/$\mu$m) & $\hat{\Phi}_\mathrm{max}$ & $10^{-4} \eta_0\Phi_\mathrm{star}$ & $10^{-5} \eta_0\Phi_\mathrm{star}$ & $10^{-6} \eta_0\Phi_\mathrm{star}$   \\
\hline 
\end{tabular}
\end{center}
\end{table} 







\acknowledgments     
The authors thank Brunella Carlomagno and Olivier Absil from the University of Li\`{e}ge (ULg), Alexis Carlotti from  Institute of Planetology and Astrophysics of Grenoble (IPAG), and Christian Delacroix from Cornell University for many fruitful discussions regarding coronagraph performance and optimization. This work was supported by the Exoplanet Exploration Program (ExEP), Jet Propulsion Laboratory, California Institute of Technology, under contract to the National Aeronautics and Space Administration (NASA).


\bibliography{RuaneLibrary}   

\begin{thebibliography}{10}

\bibitem{GPI2006}
Macintosh, B., Graham, J., Palmer, D., Doyon, R., Gavel, D., Larkin, J.,
  Oppenheimer, B., Saddlemyer, L., Wallace, J.~K., Bauman, B., Evans, J.,
  Erikson, D., Morzinski, K., Phillion, D., Poyneer, L., Sivaramakrishnan, A.,
  Soummer, R., Thibault, S., and Veran, J.-P., ``{The Gemini Planet Imager},''
  {\em Proc. SPIE}~{\bf 6272},  62720L (2006).

\bibitem{SPHERE2008}
Beuzit, J.-L., Feldt, M., Dohlen, K., Mouillet, D., Puget, P., Wildi, F., Abe,
  L., Antichi, J., Baruffolo, A., Baudoz, P., Boccaletti, A., Carbillet, M.,
  Charton, J., Claudi, R., Downing, M., Fabron, C., Feautrier, P., Fedrigo, E.,
  Fusco, T., Gach, J.-L., Gratton, R., Henning, T., Hubin, N., Joos, F.,
  Kasper, M., Langlois, M., Lenzen, R., Moutou, C., Pavlov, A., Petit, C.,
  Pragt, J., Rabou, P., Rigal, F., Roelfsema, R., Rousset, G., Saisse, M.,
  Schmid, H.-M., Stadler, E., Thalmann, C., Turatto, M., Udry, S., Vakili, F.,
  and Waters, R., ``{SPHERE: a `Planet Finder' instrument for the VLT},'' {\em
  Proc. SPIE}~{\bf 7014},  701418 (2008).

\bibitem{Martinache2009}
{Martinache}, F. and {Guyon}, O., ``{The Subaru Coronagraphic Extreme-AO
  Project},'' {\em Proc. SPIE}~{\bf 7440},  74400O (2009).

\bibitem{Marois2008}
Marois, C., Macintosh, B., Barman, T., Zuckerman, B., Song, I., Patience, J.,
  Lafrenière, D., and Doyon, R., ``{Direct Imaging of Multiple Planets
  Orbiting the Star HR 8799},'' {\em Science}~{\bf 322}(5906),  1348--1352
  (2008).

\bibitem{Lafreniere2008}
Lafreni\`{e}re, D., Jayawardhana, R., and van Kerkwijk, M.~H., ``Direct imaging
  and spectroscopy of a planetary-mass candidate companion to a young solar
  analog,'' {\em Astrophys. J.}~{\bf 689}(2),  L153--L156 (2008).

\bibitem{Lagrange2009}
Lagrange, A.-M., Gratadour, D., Chauvin, G., Fusco, T., Ehrenreich, D.,
  Mouillet, D., Rousset, G., Rouan, D., Allard, F., Gendron, E., Charton, J.,
  Mugnier, L., Rabou, P., Montri, J., and Lacombe, F., ``{A probable giant
  planet imaged in the $\beta$ Pictoris disk},'' {\em Astron. Astrophys.}~{\bf
  493}(2),  L21--L25 (2009).

\bibitem{Macintosh2015}
Macintosh, B., Graham, J.~R., Barman, T., De~Rosa, R.~J., Konopacky, Q.,
  Marley, M.~S., Marois, C., Nielsen, E.~L., Pueyo, L., Rajan, A., Rameau, J.,
  Saumon, D., Wang, J.~J., Ammons, M., Arriaga, P., Artigau, E., Beckwith, S.,
  Brewster, J., Bruzzone, S., Bulger, J., Burningham, B., Burrows, A.~S., Chen,
  C., Chiang, E., Chilcote, J.~K., Dawson, R.~I., Dong, R., Doyon, R., Draper,
  Z.~H., Duchêne, G., Esposito, T.~M., Fabrycky, D., Fitzgerald, M.~P.,
  Follette, K.~B., Fortney, J.~J., Gerard, B., Goodsell, S., Greenbaum, A.~Z.,
  Hibon, P., Hinkley, S., Cotton, T.~H., Hung, L.-W., Ingraham, P.,
  Johnson-Groh, M., Kalas, P., Lafreniere, D., Larkin, J.~E., Lee, J., Line,
  M., Long, D., Maire, J., Marchis, F., Matthews, B.~C., Max, C.~E., Metchev,
  S., Millar-Blanchaer, M.~A., Mittal, T., Morley, C.~V., Morzinski, K.~M.,
  Murray-Clay, R., Oppenheimer, R., Palmer, D.~W., Patel, R., Patience, J.,
  Perrin, M.~D., Poyneer, L.~A., Rafikov, R.~R., Rantakyrö, F.~T., Rice, E.,
  Rojo, P., Rudy, A., Ruffio, J.-B., Ruiz, M.~T., Sadakuni, N., Saddlemyer, L.,
  Salama, M., Savransky, D., Schneider, A.~C., Sivaramakrishnan, A., Song, I.,
  Soummer, R., Thomas, S., Vasisht, G., Wallace, J.~K., Ward-Duong, K.,
  Wiktorowicz, S.~J., Wolff, S.~G., and Zuckerman, B., ``{Discovery and
  spectroscopy of the young Jovian planet 51 Eri b with the Gemini Planet
  Imager},'' {\em Science}~{\bf 350}(6256),  64--67 (2015).

\bibitem{Bowler2016}
{Bowler}, B.~P., ``{Imaging Extrasolar Giant Planets},'' {\em Publ. Astron.
  Soc. Pac.}  (2016).

\bibitem{Macintosh2006}
{Macintosh}, B., {Troy}, M., {Doyon}, R., {Graham}, J., {Baker}, K., {Bauman},
  B., {Marois}, C., {Palmer}, D., {Phillion}, D., {Poyneer}, L., {Crossfield},
  I., {Dumont}, P., {Levine}, B.~M., {Shao}, M., {Serabyn}, G., {Shelton}, C.,
  {Vasisht}, G., {Wallace}, J.~K., {Lavigne}, J.-F., {Valee}, P., {Rowlands},
  N., {Tam}, K., and {Hackett}, D., ``{Extreme adaptive optics for the Thirty
  Meter Telescope},'' {\em Proc. SPIE}~{\bf 6272},  62720N (2006).

\bibitem{Kasper2010}
Kasper, M., Beuzit, J.-L., Verinaud, C., Gratton, R.~G., Kerber, F., Yaitskova,
  N., Boccaletti, A., Thatte, N., Schmid, H.~M., Keller, C., Baudoz, P., Abe,
  L., Aller-Carpentier, E., Antichi, J., Bonavita, M., Dohlen, K., Fedrigo, E.,
  Hanenburg, H., Hubin, N., Jager, R., Korkiakoski, V., Martinez, P., Mesa, D.,
  Preis, O., Rabou, P., Roelfsema, R., Salter, G., Tecza, M., and Venema, L.,
  ``{EPICS: direct imaging of exoplanets with the E-ELT},'' {\em Proc.
  SPIE}~{\bf 7735},  77352E (2010).

\bibitem{Noecker2016}
Noecker, M.~C., Zhao, F., Demers, R., Trauger, J., Guyon, O., and Kasdin,
  N.~J., ``{Coronagraph instrument for WFIRST-AFTA},'' {\em J. Astron. Telesc.
  Instrum. Syst.}~{\bf 2}(1),  011001 (2016).

\bibitem{Feinberg2014}
Feinberg, L.~D., Jones, A., Mosier, G., Rioux, N., Redding, D., and Kienlen,
  M., ``{A cost-effective and serviceable ATLAST 9.2m telescope
  architecture},'' {\em Proc. SPIE}~{\bf 9143},  914316 (2014).

\bibitem{Dalcanton2015}
{Dalcanton}, J., {Seager}, S., {Aigrain}, S., {Battel}, S., {Brandt}, N.,
  {Conroy}, C., {Feinberg}, L., {Gezari}, S., {Guyon}, O., {Harris}, W.,
  {Hirata}, C., {Mather}, J., {Postman}, M., {Redding}, D., {Schiminovich}, D.,
  {Stahl}, H.~P., and {Tumlinson}, J., ``{From Cosmic Birth to Living Earths:
  The Future of UVOIR Space Astronomy},'' {\em ArXiv e-prints} ,  1507.04779
  (2015).

\bibitem{Kuchner2002}
{Kuchner}, M.~J. and {Traub}, W.~A., ``{A Coronagraph with a Band-limited Mask
  for Finding Terrestrial Planets},'' {\em Astrophys. J.}~{\bf 570},  900--908
  (2002).

\bibitem{Kasdin2003}
{Kasdin}, N.~J., {Vanderbei}, R.~J., {Spergel}, D.~N., and {Littman}, M.~G.,
  ``{Extrasolar Planet Finding via Optimal Apodized-Pupil and Shaped-Pupil
  Coronagraphs},'' {\em Astrophys. J.}~{\bf 582},  1147--1161 (2003).

\bibitem{Codona2004}
{Codona}, J.~L. and {Angel}, R., ``Imaging extrasolar planets by stellar halo
  suppression in separately corrected color bands,'' {\em Astrophys. J.}~{\bf
  604}(2),  L117 (2004).

\bibitem{Mawet2005}
{Mawet}, D., {Riaud}, P., {Absil}, O., and {Surdej}, J., ``{Annular Groove
  Phase Mask Coronagraph},'' {\em Astrophys. J.}~{\bf 633},  1191--1200 (2005).

\bibitem{Foo2005}
{Foo}, G., {Palacios}, D.~M., and {Swartzlander}, G.~A., ``{Optical vortex
  coronagraph},'' {\em Opt. Lett.}~{\bf 30},  3308--3310 (2005).

\bibitem{Guyon2005}
{Guyon}, O., {Pluzhnik}, E.~A., {Galicher}, R., {Martinache}, F., {Ridgway},
  S.~T., and {Woodruff}, R.~A., ``{Exoplanet Imaging with a Phase-induced
  Amplitude Apodization Coronagraph. I. Principle},'' {\em Astrophys. J.}~{\bf
  622},  744--758 (2005).

\bibitem{Soummer2005}
Soummer, R., ``{Apodized Pupil Lyot Coronagraphs for Arbitrary Telescope
  Apertures},'' {\em Astrophys. J.}~{\bf 618}(2),  L161 (2005).

\bibitem{Trauger2007}
{Trauger}, J.~T. and {Traub}, W.~A., ``{A laboratory demonstration of the
  capability to image an Earth-like extrasolar planet},'' {\em Nature}~{\bf
  446},  771--773 (2007).

\bibitem{Guyon2010}
{Guyon}, O., {Martinache}, F., {Belikov}, R., and {Soummer}, R., ``{High
  Performance PIAA Coronagraphy with Complex Amplitude Focal Plane Masks},''
  {\em Astrophys. J. Suppl. Ser.}~{\bf 190},  220--232 (2010).

\bibitem{Mawet2011_improved}
{Mawet}, D., {Serabyn}, E., {Wallace}, J.~K., and {Pueyo}, L., ``{Improved
  high-contrast imaging with on-axis telescopes using a multistage vortex
  coronagraph},'' {\em Opt. Lett.}~{\bf 36},  1506 (2011).

\bibitem{Mawet2013_ringapod}
{Mawet}, D., {Pueyo}, L., {Carlotti}, A., {Mennesson}, B., {Serabyn}, E., and
  {Wallace}, J.~K., ``{Ring-apodized Vortex Coronagraphs for Obscured
  Telescopes. I. Transmissive Ring Apodizers},'' {\em Astrophys. J. Suppl.
  Ser.}~{\bf 209},  7 (2013).

\bibitem{Carlotti2014}
{Carlotti}, A., {Pueyo}, L., and {Mawet}, D., ``{Apodized phase mask
  coronagraphs for arbitrary apertures. II. Comprehensive review of solutions
  for the vortex coronagraph},'' {\em Astron. Astrophys.}~{\bf 566},  A31
  (2014).

\bibitem{Ruane2015_SPIE}
Ruane, G.~J., Absil, O., Huby, E., Mawet, D., Delacroix, C., Carlomagno, B.,
  Piron, P., and Swartzlander, G.~A., ``{Optimized focal and pupil plane masks
  for vortex coronagraphs on telescopes with obstructed apertures},'' {\em
  Proc. SPIE}~{\bf 9605},  96051I (2015).

\bibitem{Ruane2015_LPM}
Ruane, G.~J., Huby, E., Absil, O., Mawet, D., Delacroix, C., Carlomagno, B.,
  and Swartzlander, G.~A., ``{Lyot-plane phase masks for improved high-contrast
  imaging with a vortex coronagraph},'' {\em A\&A}~{\bf 583},  A81 (2015).

\bibitem{Pueyo2013}
{Pueyo}, L. and {Norman}, C., ``{High-contrast Imaging with an Arbitrary
  Aperture: Active Compensation of Aperture Discontinuities},'' {\em Astrophys.
  J.}~{\bf 769},  102 (2013).

\bibitem{Mazoyer2015}
Mazoyer, J., Pueyo, L., Norman, C., N'Diaye, M., Mawet, D., Soummer, R.,
  Perrin, M., Choquet, E., and Carlotti, A., ``{Active correction of aperture
  discontinuities (ACAD) for space telescope pupils: a parametic analysis},''
  {\em Proc. SPIE}~{\bf 9605},  96050M (2015).

\bibitem{Guyon2014}
{Guyon}, O., {Hinz}, P.~M., {Cady}, E., {Belikov}, R., and {Martinache}, F.,
  ``{High Performance Lyot and PIAA Coronagraphy for Arbitrarily Shaped
  Telescope Apertures},'' {\em Astrophys. J.}~{\bf 780},  171 (2014).

\bibitem{Balasubramanian2015}
Balasubramanian, K., White, V., Yee, K., Echternach, P., Muller, R., Dickie,
  M., Cady, E., Prada, C.~M., Ryan, D., Poberezhskiy, I., Kern, B., Zhou, H.,
  Krist, J., Nemati, B., Eldorado~Riggs, A.~J., Zimmerman, N.~T., and Kasdin,
  N.~J., ``{WFIRST-AFTA coronagraph shaped pupil masks: design, fabrication,
  and characterization},'' {\em J. Astron. Telesc. Instrum. Syst.}~{\bf 2}(1),
  011005 (2015).

\bibitem{Trauger2016}
Trauger, J., Moody, D., Krist, J., and Gordon, B., ``{Hybrid Lyot coronagraph
  for WFIRST-AFTA: coronagraph design and performance metrics},'' {\em J.
  Astron. Telesc. Instrum. Syst.}~{\bf 2}(1),  011013 (2016).

\bibitem{Zimmerman2016}
Zimmerman, N.~T., Eldorado~Riggs, A.~J., Kasdin, N.~J., Carlotti, A., and
  Vanderbei, R.~J., ``{Shaped pupil Lyot coronagraphs: high-contrast solutions
  for restricted focal planes},'' {\em J. Astron. Telesc. Instrum. Syst.}~{\bf
  2}(1),  011012 (2016).

\bibitem{Serabyn2010}
{Serabyn}, E., {Mawet}, D., and {Burruss}, R., ``{An image of an exoplanet
  separated by two diffraction beamwidths from a star},'' {\em Nature}~{\bf
  464},  1018--1020 (2010).

\bibitem{Mawet2010b}
{Mawet}, D., {Pueyo}, L., {Moody}, D., {Krist}, J., and {Serabyn}, E., ``{The
  Vector Vortex Coronagraph: sensitivity to central obscuration, low-order
  aberrations, chromaticism, and polarization},'' {\em Proc. SPIE}~{\bf 7739},
  14 (2010).

\bibitem{SCDA}
{Feinberg}, L., {Hull}, T., {Knight}, J.~S., {Krist}, J., {Lightsey}, P.,
  {Matthews}, G., Shaklan, S., and {Stahl}, H.~P., ``Apertures for segmented
  coronagraph design and analysis.''
  \url{http://exep.jpl.nasa.gov/files/exep/SCDAApertureDocument0504161.pdf}.

\bibitem{Soummer2007}
Soummer, R., Ferrari, A., Aime, C., and Jolissaint, L., ``Speckle noise and
  dynamic range in coronagraphic images,'' {\em Astrophys. J.}~{\bf 669}(1),
  642 (2007).

\bibitem{Martinache2014}
Martinache, F., Guyon, O., Jovanovic, N., Clergeon, C., Singh, G., Kudo, T.,
  Currie, T., Thalmann, C., McElwain, M., and Tamura, M., ``{On-Sky Speckle
  Nulling Demonstration at Small Angular Separation with SCExAO},'' {\em Publ.
  Astron. Soc. Pac.}~{\bf 126}(940),  565 (2014).

\bibitem{Marois2006}
{Marois}, C., {Lafreni{\`e}re}, D., {Doyon}, R., {Macintosh}, B., and {Nadeau},
  D., ``{Angular Differential Imaging: A Powerful High-Contrast Imaging
  Technique},'' {\em Astrophys. J.}~{\bf 641},  556--564 (2006).

\bibitem{Soummer2012}
{Soummer}, R., {Pueyo}, L., and {Larkin}, J., ``{Detection and Characterization
  of Exoplanets and Disks Using Projections on Karhunen-Lo{\`e}ve
  Eigenimages},'' {\em Astrophys. J. Lett.}~{\bf 755},  L28 (2012).

\bibitem{Mawet2009}
{Mawet}, D., {Serabyn}, E., {Liewer}, K., {Hanot}, C., {McEldowney}, S.,
  {Shemo}, D., and {O'Brien}, N., ``{Optical Vectorial Vortex Coronagraphs
  using Liquid Crystal Polymers: theory, manufacturing and laboratory
  demonstration},'' {\em Opt. Express}~{\bf 17},  1902--1918 (2009).

\bibitem{Delacroix2013}
{Delacroix}, C., {Absil}, O., {Forsberg}, P., {Mawet}, D., {Christiaens}, V.,
  {Karlsson}, M., {Boccaletti}, A., {Baudoz}, P., {Kuittinen}, M.,
  {Vartiainen}, I., {Surdej}, J., and {Habraken}, S., ``{Laboratory
  demonstration of a mid-infrared AGPM vector vortex coronagraph},'' {\em
  Astron. Astrophys.}~{\bf 553},  A98 (2013).

\bibitem{Mawet2014}
Mawet, D., Shelton, C., Wallace, J., Bottom, M., Kuhn, J., Mennesson, B.,
  Burruss, R., Bartos, R., Pueyo, L., Carlotti, A., and Serabyn, E., ``{
  Demonstration of vortex coronagraph concepts for on-axis telescopes on the
  Palomar Stellar Double Coronagraph },'' {\em Proc. SPIE}~{\bf 9143},  91432T
  (2014).

\end{thebibliography}
\bibliographystyle{spiebib}   

\end{document}